\documentclass{aa}

\usepackage{epsfig}
\newcommand{\msun}{\mbox{$M_{\odot}$}}
\newcommand{\Msun}{\mbox{$M_{\odot}$}}
\newcommand{\lsun}{\mbox{$L_{\odot}$}}

\newcommand{\zsun}{\mbox{$Z_{\odot}$}}
\newcommand{\teff}{\mbox{$T_{\rm eff}$}}
\newcommand{\Teff}{\mbox{$T_{\rm eff}$}}

\newcommand{\vinf}{\mbox{$v_{\infty}$}}
\newcommand{\vesc}{\mbox{$v_{\rm esc}$}}
\newcommand{\ratio}{\mbox{$v_{\infty}$/$v_{\rm esc}$}}
\newcommand{\mdot}{\mbox{$\dot{M}$}}

\newcommand{\msunyr}{\mbox{$M_{\odot} {\rm yr}^{-1}$}}

\begin{document}

\thesaurus{07(08.05.1; 08.13.2; 08.19.3; 08.23.3; 08.05.3)}

\title{Mass-loss predictions for O and B stars as a function of metallicity}

\author{Jorick S. Vink\inst{1}
 \and Alex de Koter\inst{2}
 \and Henny J.G.L.M. Lamers\inst{1,3}}
\offprints{Jorick S. Vink, j.vink@ic.ac.uk}

\institute{ Astronomical Institute, Utrecht University,
            P.O.Box 80000, NL-3508 TA Utrecht, The Netherlands
            \and
            Astronomical Institute ``Anton Pannekoek'', University of Amsterdam,
            Kruislaan 403, NL-1098 SJ Amsterdam, The Netherlands
            \and
            SRON Laboratory for Space Research, Sorbonnelaan 2, NL-3584 CA Utrecht, The Netherlands}

\titlerunning{Predicted $\dot{M}$ for different $Z$}
\authorrunning{Jorick S. Vink et al.}

\maketitle
\begin{abstract}

We have calculated a grid of massive star wind models and 
mass-loss rates for a wide range of metal abundances between 
$1/100 \le Z/\zsun \le 10$. 

The calculation of this grid completes 
the Vink et al. (2000) mass-loss recipe with an additional 
parameter $Z$.
We have found that the exponent of the power law dependence of mass loss 
vs. metallicity is constant in the range between 1/30 $\le$ $Z/\zsun$ $\le$ 3. 
The mass-loss rate scales as $\dot{M} \propto Z^{0.85} \vinf^p$
with $p = -1.23$ for stars with $\teff \ga 25~000$ K, and $p = -1.60$ for 
the B supergiants with $\teff \la 25~000$ K. Taking also into account the metallicity 
dependence of $\vinf$, using the power law dependence $\vinf \propto Z^{0.13}$ from Leitherer et al. (1992), 
the overall result of mass loss as a function of metallicity can be represented by 
$\dot{M} \propto Z^{0.69}$ for stars with $\teff \ga 25~000$ K, and $\dot{M} \propto Z^{0.64}$ 
for B supergiants with $\teff \la 25~000$ K.

Although it is derived that the exponent of the mass loss vs. metallicity
dependence is constant over a large range in $Z$, one should 
be aware of the presence of bi-stability jumps at specific 
temperatures. Here the character of the line driving changes 
drastically due to recombinations of dominant metal species 
resulting in jumps in the mass loss.
We have investigated the physical origins of these jumps and have 
derived formulae that combine mass loss recipes for 
both sides of such jumps. As observations of different galaxies 
show that the ratio Fe/O varies with metallicity, we make a distinction 
between the metal abundance $Z$ derived on the basis of iron or oxygen
lines.

Our mass-loss predictions are successful in explaining 
the observed mass-loss rates for Galactic and Small Magellanic Cloud 
O-type stars, as well as in predicting the observed Galactic 
bi-stability jump. Hence, we believe that our predictions are reliable 
and suggest that our mass-loss recipe be used in future 
evolutionary calculations of massive stars at different metal
abundance.
A computer routine to calculate mass loss is publicly 
available.


\end{abstract}


\section{Introduction}
\label{s_intro}

In this paper we predict the rate at which mass is lost due to stellar 
winds from massive O and B-type stars as a function of metal 
abundance: ${\dot{M} = f(Z)}$.  The calculations are based on 
state-of-the-art modelling. The model description takes into account 
momentum transfer of radiation to gas in a way that photons 
are allowed to interact with ions in the wind more than just once. 
This method, which was pioneered by Abbott \& Lucy (1985) and Schmutz et al. (1991),
has been used in a previous study (Vink et al. 2000) where wind 
models including the effects of ``multiple scattering'' were calculated for Galactic 
early-type stars. It was shown that our predictions agree with the 
observations for Galactic O stars, which resolved a persistent discrepancy between 
observed and theoretical mass-loss rates (Lamers \& Leitherer 1993, 
Puls et al. 1996).

Metallicity is a key parameter that controls many aspects of 
the formation and the evolution of both stars and galaxies.
For instance, the overall chemical enrichment of the interstellar medium (ISM) 
is a strong function of metallicity. Secondly, the relative importance of 
stellar winds compared to Supernova explosions depends on $Z$ in the sense that 
stellar winds become more important with increasing metallicity 
(Leitherer et al. 1992). Since chemical elements are produced in stars 
with different masses, they enrich the ISM on different timescales. 
Massive stars mainly contribute to the enrichment of oxygen, other $\alpha$-elements 
and iron. Therefore, these elements are ejected on short timescales. 
Although carbon and nitrogen are also produced in massive stars,  
their main contribution comes from longer-lived 
intermediate mass stars. This implies that if the star formation history 
and the initial mass function are considered, metallicity is expected to cause 
a ``differential'' chemical enrichment of the ISM in different galaxies.

Recent models of the chemical evolution versus redshift in the Universe 
predict that metallicity shows a stronger dependence on the local density 
(i.e. galaxy mass) than on redshift (Cen \& Ostriker 1999). Hence, galaxies with 
high and low metal abundances are expected to be found at all cosmological distances. 
These models reasonably predict the range in metal abundance that has been observed. 
The metallicity reaches as high as 10 times the solar value $Z_{\odot}$ in central 
regions of active galactic nuclei and quasars (Artymowicz 1993, Hamann 1997), but is 
only about 1/50 $Z_{\odot}$ for the blue compact dwarf galaxy IZw18 
(Sargent \& Searle 1970, Izotov \& Thuan 1999). Such low metallicity may 
imply that blue compact dwarf galaxies only experience their first episode 
of star formation.
Based on the observed range in $Z$, we will study the mass loss properties of massive 
stars within the representative 
metallicity range of $1/100 \leq Z/Z_{\odot} \leq 10$. 
     
The driving mechanism of the winds of massive early-type stars
is radiation pressure on numerous spectral lines (Castor et al. 1975, 
hereafter CAK; Abbott 1982, Pauldrach et al. 1986, Vink et al. 2000).
It is important to know {\em which} lines are actually
responsible for the acceleration of the winds. As hydrogen and helium 
only have very few lines in the relevant spectral range in which early-type 
stars emit most of their radiation, it is mainly lines of the {\it metals} that 
are responsible for the line driving. This thus implies that the stellar 
wind strengths are expected to depend on metal abundance.

Observational evidence for metallicity dependent
stellar wind properties was found by Garmany \& Conti (1985) and Prinja (1987). 
They found that the terminal flow velocity of the stellar wind in the Magellanic Cloud
stars was lower than that of Galactic stars. The authors attributed this 
difference to an under-abundance of metals in the Magellanic Clouds
compared to the Galaxy. 

The quantitative dependence of $\dot{M}$ on $Z$ was theoretically 
studied by CAK, Abbott (1982) and Kudritzki et al. (1987). These 
studies have shown that the $\mdot(Z)$ relation is expected
to behave as a power-law:

\begin{equation}
\mdot \propto Z^{m}
\label{eq_mdotz}
\end{equation}
with predictions for the index $m$ ranging between about 1/2 (Kudritzki et al. 1987) 
to 0.94 (Abbott 1982). 
Since these results were based on radiation-driven wind models that did not 
take into account the effect of ``multiple scattering'', a new investigation of 
the $\dot{M}$ vs. $Z$ relation, is appropriate. Especially since Eq.~(\ref{eq_mdotz}) is 
widely used in evolutionary calculations for massive stars, usually adopting $m = 1/2$ 
(e.g. Meynet et al. 1994). 

We will use our ``Unified Monte Carlo'' method (Vink et al. 2000) 
to predict mass-loss rates of early-type stars over a wide range in 
metallicities and stellar parameters. 
In this approach, multiple scatterings are consistently taken into account 
and an artificial separation between the 
stellar photosphere and wind (core-halo) is avoided. The main
question we will address is: 'What is the dependence of stellar mass
loss on metal abundance ?'.

In Sects.~\ref{s_method} and~\ref{s_assumptions}, the method to 
calculate mass-loss rates and the adopted assumptions will 
be described. 
In Sect.~\ref{s_massloss}, the resulting 
wind models and mass-loss rates will be presented. The relative 
importance of Fe and CNO elements to the line force will be 
discussed in Sect.~\ref{s_relabund}.
In Sects.~\ref{s_metal} and~\ref{s_recipe} the dependence of the mass-loss rate
on metallicity will be determined. This completes 
the Vink et al. (2000) mass-loss recipe to predict $\dot{M}$ 
as a function of stellar parameters with an additional $Z$ dependence.
It will be shown that over a large parameter space, 
the exponent of the $\dot{M}(Z)$ power law dependence is constant, 
but that at specific temperatures, one needs to take the presence of so-called 
bi-stability jumps into account.
In Sect.~\ref{s_comp} these  
mass-loss predictions will be compared with observed mass-loss rates
for the Large Magellanic Cloud and the Small Magellanic Cloud. 
Finally, in Sect.~\ref{s_concl}, the study will be summarised.


\section{Theoretical context}
\label{s_context}

In this section we will discuss the basic physical processes that may play 
a role in determining the dependence of mass loss on metal abundance.
We will describe the expected effects in terms of CAK theory. However,
in our detailed predictions (Sect.~\ref{s_massloss}), we will not use this 
formalism, but extend on it by including multiple scattering effects.

In CAK theory the line acceleration is conveniently expressed in units of the
force multiplier $M(t)$ and is given by (CAK, Abbott 1982):
 
\begin{equation}
M(t)~=~k~t^{- \alpha}~\left(\frac{n_{\rm e}}{W}\right)^\delta
\label{eq_force}
\end{equation}
where $n_{\rm e}$ is the electron density and $W$ is the geometrical
dilution factor. The parameters $k$, $\alpha$ and $\delta$ are the 
so-called force multiplier parameters. The first one, $k$, 
is a measure for the number of lines. The second parameter, $\alpha$, 
describes the ratio of the optically thick line acceleration over 
the total line acceleration (Gayley 1995, Puls et al. 2000)
If only strong (weak) lines contribute to the force, then
$\alpha$ = 1 (0). The predicted value of $\alpha$ for O-type stars is
typically 0.6 (Abbott 1982, Kudritzki et al. 1989). The parameter
$\delta$ describes the ionization in the wind. Its value is 
usually $\delta \sim 0.1$. Finally, $t$ is the optical depth parameter, given by:

\begin{equation}
t~=~ \sigma_e v_{\rm th} \rho (dr/dv)
\label{eq_tCAK}
\end{equation}
where $v_{th}$ is the mean thermal velocity of the protons
and $\sigma_e$ is the electron scattering cross-section.

Abbott (1982) and Puls et al. (2000) have shown
that the CAK force-multiplier parameter $k$ is
dependent on the metallicity in the following way:

\begin{equation}
k(Z)~\propto~Z^{1 - \alpha}
\label{eq_kprop}
\end{equation}
Kudritzki et al. (1989) have calculated analytical
solutions for radiation-driven wind models that include
the finite cone angle effect. The scaling relation for the mass-loss
rate that was derived, is proportional to

\begin{equation}
\dot{M}~\propto~k^{1/\alpha_{\rm eff}}
\label{eq_mdotprop}
\end{equation}
where 
\begin{equation}
\alpha_{\rm eff} = \alpha - \delta
\label{eq_alphaeff}
\end{equation} 
This implies that $\dot{M}$ is expected
to depend on metallicity in the following way:

\begin{equation}
\dot{M}~\propto~Z^m~
\label{eq_mdot}
\end{equation}
with

\begin{equation}
m~=~\frac{1 - \alpha}{\alpha - \delta}
\label{eq_md}
\end{equation}
Since a typical value for $m$ is (1 - 0.6)/(0.6 - 0.1) = 0.8, one  
would expect a more linear ($m \simeq 0.8$)
dependence of $\dot{M}$ on $Z$, instead of the
square-root ($m$ = 1/2) dependence that was
calculated by Kudritzki et al. (1987). 
However, as the force multiplier parameter $\alpha$ itself is 
dependent on metallicity, the terminal velocity $\vinf$ also becomes 
a function of $Z$: $\vinf \propto Z^q$. Note that Leitherer et al. (1992)
indeed derived such a more  linear ($m \simeq 0.8$)
dependence of $\dot{M}$ on $Z$, and additionally derived $\vinf \propto Z^{0.13}$. 
However, multi-line transfer was not taken into account in these calculations either. 

We note that a pure power-law dependence 
of $\dot{M}$ on $Z$ over the entire parameter space, is 
questionable.  It may be expected that for a certain 
metallicity range Eq.~(\ref{eq_mdotz}) provides a useful representation 
of the mass loss vs. metallicity relation, but that at some 
minimum and maximum $Z$, deviations from a power-law may occur. 
For instance, deviations at high metallicity may occur when mass 
loss is so efficient that densities in the wind are so high that all 
relevant Fe lines become saturated. Hence, at some point, an increase in 
metallicity may no longer cause a substantial increase in mass loss and 
subsequently a flattening of the $\dot{M}(Z)$ relation is expected. 
Deviations at low metallicity, with subsequently low mass loss, 
may occur when only weak iron lines remain present. 
Other abundant ions, such as those of C, N, and O, which normally have far fewer effective driving 
lines than Fe, may start to dominate the driving because their main lines 
are still strong. Again a shallower slope of the $\mdot(Z)$ relation 
is anticipated.

A second important item in the calculations of mass loss at 
different $Z$, is the possible presence of one or more ``bi-stability'' 
jumps at different $Z$. For Galactic metallicities, at an effective 
temperature of $\sim 25 000$ K, the mass loss is predicted to increase 
dramatically by a factor of about five. The effect of this jump
on terminal velocity has observationally 
been found by Lamers et al. (1995). The origin for this jump is 
related to a recombination from Fe {\sc iv} to {\sc iii} in the lower
part of the wind (Vink et al. 1999).
Since the ionization equilibrium does not only depend 
on temperature, but also on density, one may
expect a shift in the position of this bi-stability 
jump as a function of $Z$. 
Moreover, at lower metallicity, other abundant ions, such as 
those of CNO, may start to dominate the wind driving, 
implying there could be additional bi-stability jumps at 
different $Z$ due to recombinations of one of these elements.

In this paper we will therefore concentrate on three main
issues: firstly, the global dependence of the mass-loss rate
on $Z$; secondly, the presence and position of bi-stability 
jumps for different $Z$, and, thirdly, the relative importance
of Fe and CNO elements at low metal abundance.

\section{Method to calculate \mdot}
\label{s_method}

The mass-loss rates are calculated with a Monte Carlo (MC)
method that follows the fate of a large number of
photons from below the stellar photosphere through the wind
and calculates the radiative acceleration of the wind
material.
The core of the approach is that the total loss of 
radiative energy is coupled to the momentum gain 
of the outflowing material.
Since the absorptions and scatterings of photons in the wind
depend on the density in the wind and hence on the mass-loss
rate, it is possible to find a consistent model where the momentum
of the wind material is exactly equal to the radiative momentum
that has been transferred.
The method is similar to the technique introduced 
by Abbott \& Lucy (1985). The precise characteristics of our 
Unified MC approach have been described in Vink et al. (1999).
The essential ingredients and the assumptions of our 
approach have extensively been discussed in Vink et al. (2000).
 
The MC code uses a density and temperature structure that
has been computed in a prior model atmosphere calculation ({\sc isa-wind}).
The model atmospheres used for this study are calculated
with the non-LTE unified Improved Sobolev Approximation
code ({\sc isa-wind}) for stars with extended 
atmospheres. For details of the model atmosphere we refer the reader 
to de Koter et al. (1993, 1997). The chemical species that 
are explicitly calculated in non-LTE are H, He, C, N, O 
and Si. The iron-group elements, which are important for 
the radiative driving and consequently for $\dot{M}$, are
treated in a generalised version of the 
``modified nebular approximation'' (Schmutz 1991).

The temperature structure of the {\sc isa-wind} model 
atmosphere is based on the grey LTE approximation. This 
implies that radiative equilibrium is not strictly fulfilled, but that 
deviations at the one percent level may occur in {\sc isa-wind}. 
In contrast, local radiative equilibrium is automatically enforced in {\sc mc-wind}. We mention that the total opacity that is treated in {\sc mc-wind} is 
larger than that treated in {\sc isa-wind}. 
Regarding emissions, the frequency distribution of thermally emitted photons in {\sc mc-wind} 
is only based on the elements that were explicitly computed in the {\sc isa-wind} atmosphere calculation. Regarding absorptions, the MC simulations 
include those due to metal ions (mostly iron), 
which are not accounted for in {\sc isa-wind}. This inconsistency may 
introduce a small discrepancy in the frequency distribution between true emission and 
true absorption, causing an underestimate of thermal emissions relative 
to absorptions in spectral regions of high iron opacity, whereas in all
other regions of the spectral energy distribution the situation is reversed. 
Nevertheless, because {\sc mc-wind} conserves total energy, we do not 
expect this effect to influence the predicted mass-loss rates significantly.

The line list that is used for these MC calculations consists of 
over $10^5$ of the strongest transitions of the elements H~-~Zn 
extracted from 
the line list constructed by Kurucz (1988). Lines in the wavelength 
region between 50 and 7000 \AA~are included in the calculations 
with ionization stages up to stage {\sc vi}.
The wind was divided into 
about 50-60 concentric shells, with many narrow shells in the subsonic 
region and wider shells in supersonic layers. 
The division in shells is essentially made on the basis 
of the Rosseland optical depth scale, with typical changes in the 
logarithm of the optical depth of about 0.13. 
For each set of model parameters a certain number
of photon packets is followed.
For Galactic metallicities this number is typically 
about $2~10^5$ (see Vink et al. 2000) 

At lower $Z$, and consequently at lower mass-loss rates, however, 
the typical amount of photon packets has to be increased, to keep up good 
statistics, as one is shooting photons through a 
less dense wind. Consequently, photon packets experience smaller numbers 
of line interactions. We found that as long as there were typically 
$\sim$ 100 line scatterings in each supersonic shell, the derived
mass loss was reasonably accurate, i.e. $\Delta$ log $\dot{M} \la 0.05$.  

At extremely low metallicities ($Z/\zsun \la 1/30$) the 
line driving mechanism becomes very inefficient and accurate wind solutions
can only be obtained for the highest stellar luminosities, i.e. 
log $L/\lsun \ga$ 6. Hence, the lowest $Z$ models ($Z/\zsun$ = 1/100) 
will only be calculated for $L/\lsun = 6$ (see Sect.~\ref{s_massloss}).


\section{The assumptions of the model grid}
\label{s_assumptions}

For every $Z$, the mass-loss rate was calculated for 12 
values of $\teff$ in the range between 12~500 and 50~000 K.

The abundances of the metallicity grid are 
given in Table~\ref{t_abund}. $Z$ is the {\it total} metallicity content of all elements
heavier than helium. Throughout the paper we will indicate the absolute value of the metals 
with $Z$ and the value of metallicity relative to the Sun by $Z/\zsun$, adopting 
$\zsun = 0.019$ (Anders \& Grevesse 1989).
For every value of $Z$, the helium and hydrogen abundances, $Y$ and $X$ respectively,
need be adjusted accordingly. $X$ is simply given by

\begin{equation} 
X~=~1-~Y-~Z~
\end{equation}
For $Y$ we adjust the abundances in 
the following way

\begin{equation} 
Y~=~Y_{p}~+\left(\frac{\Delta Y}{\Delta Z}\right) Z
\label{eq_he}
\end{equation}
where $Y_{p}$ is the primordial helium abundance 
and ($\Delta Y/\Delta Z$) is an observed constant,
discussed below.

\begin{table}
\caption[]{Adopted abundances of the wind models.}
\label{t_abund}
\begin{tabular}{clll}
\hline
\noalign{\smallskip}
($Z/\zsun$) & $X$  & $Y$ & $Z$ \\
\noalign{\smallskip}
\hline
1/30 & 0.758 & 0.242 & 0.00063 \\
1/10  & 0.752 & 0.246 & 0.0019 \\
1/3  & 0.733 & 0.260 & 0.0063 \\
1    & 0.68  & 0.30  & 0.019 \\
3    & 0.52  & 0.42  & 0.057 \\
\noalign{\smallskip}
\hline
\end{tabular}
\end{table}

We enumerate the assumptions in the model grid:

\begin{enumerate}

\item{} Following Schaller et al. (1992) we adopt a primordial
helium abundance of $Y_{\rm p}$ = 0.24 (Audouze 1987) and 
a ($\Delta Y/\Delta Z$) ratio of 3 (Pagel 1992). The scaled 
solar metallicities were take from Allen (1973). 

\item{} All models have effective temperatures between 12~500 and 50~000 K 
with a stepsize of 2~500 K in the range 12~500 - 30~000 K 
and a stepsize of 5~000 K for the range between 30~000 and 50~000 K. 

\item{} To investigate whether the dependence of $\dot{M}$ on $Z$ is universal
for different luminosity and mass, it is 
calculated for three different values of the Eddington factor $\Gamma_e$.
This is the ratio between the gravitational and radiative acceleration 
due to electron scattering and is given by:

\begin{equation}
\label{eq_gammae}
\Gamma_e~=~\frac{L \sigma_e}{4 \pi c G M}~=~7.66~10^{-5} \sigma_{e} 
\left(\frac{L}{\lsun}\right) \left(\frac{M}{\msun}\right)^{-1}
\end{equation}
where $\sigma_e$ is the electron scattering cross-section per unit mass (its dependence 
on $\teff$ and composition is described in Lamers \& Leitherer 1993). The 
other constants have their usual meaning. The values for $\Gamma_e$ 
are given in column 1 of Table~\ref{t_parameters}.
The corresponding luminosities and masses are given in
columns 2 and 3 of the same table.

\item{} Also the dependence of $\mdot$ on the adopted ratio of the terminal flow 
velocity over the escape velocity, $\ratio$, was 
determined. Lamers et al. (1995) found that for Galactic supergiants 
the ratio $\ratio \simeq 2.6$ for stars of types
earlier than B1 and drops to $\ratio \simeq 1.3$ for stars later than type B1. 
Therefore, we have calculated mass-loss rates for input ratios of $\ratio$ of
1.3, 2.0 and 2.6 to investigate the mass loss for different values of this ratio.

We are aware that these ratios $\ratio$ may vary for different metallicity. 
However, our goal here is to determine the dependence of mass loss on different 
stellar parameters, including $\ratio$. 
If new observations with e.g. the {\it Far Ultraviolet Spectroscopic Explorer} 
show that the observed values of $\vinf$ at other $Z$ are significantly 
different from Galactic values, the predicted mass-loss rates can
easily be scaled to accommodate the new values of $\ratio$.
 
\item{} We have calculated $\dot{M}$ for wind models with a $\beta$-type 
velocity law for the accelerating part of the wind:
\begin{equation}
\label{eq_betalaw}
v(r)~=~\vinf~\left(1~-~\frac{R_*}{r}\right)^\beta
\end{equation}
Below the sonic point, a smooth transition from this 
velocity structure is made to a velocity that follows from the
photospheric density structure.
Vink et al. (2000) have shown that 
the predicted mass-loss rate is essentially insensitive to the 
adopted value of $\beta$. A value of $\beta=1$ was adopted for the 
supersonic velocity law. 

\end{enumerate}

\begin{table}
\caption[]{Adopted stellar and wind Parameters for the set of unified models.}
\label{t_parameters}
\begin{tabular}{lrcccc}
\hline
\noalign{\smallskip}
$\Gamma_{\rm e}$ & log$L_*$ & $M_*$ & \teff & ($Z/\zsun$) & $\left(\frac{\vinf}{\vesc}\right)$\\
\noalign{\smallskip}
      & ($\lsun$) & ($\Msun$) & (kK) & Range &    \\
\hline
0.130 & 5.0 &  20 & 12.5 - 50.0 & 1/30  -  3 &  1.3 - 2.6\\
0.206 & 5.5 &  40 & 12.5 - 50.0 & 1/30  -  3 &  1.3 - 2.6\\
0.434 & 6.0 &  60 & 12.5 - 50.0 & 1/100 - 10 &  1.3 - 2.6\\
\noalign{\smallskip}
\hline
\end{tabular}
\end{table}

The total grid thus contains 540 models.
Note that for {\it each} calculated point in the grid, {\it several} 
wind models had to be calculated to derive the mass-loss 
rate that is consistent with the radiative acceleration 
(see Lucy \& Abbott 1993). This results in accurate 
and self-consistent values for $\mdot$ (see Vink et al. 1999). 

\begin{table*}
\caption[]{Predicted mass-loss rates for different metallicities.}
\label{t_massloss}
\begin{tabular}{ccccrccccccc}
\hline
\noalign{\smallskip}
\multicolumn{5}{c}{} & \multicolumn{5}{c}{log $\dot{M}(\msunyr)$} \\
\noalign{\smallskip}
$\Gamma_{\rm e}$ & log$L_*$ & $M_*$ & $\ratio$ & $\teff$ & 1/100 & 1/30 & 1/10 & 1/3 & 1 & 3 & 10 \\
\noalign{\smallskip}
         & ($\lsun$) & ($\msun$) & & (kK) & $Z/\zsun$ &$Z/\zsun$ & $Z/\zsun$ & $Z/\zsun$ & $Z/\zsun$ & $Z/\zsun$ & $Z/\zsun$ \\
\hline
0.130 & 5.0 & 20 & 2.6   &  50   &  --  &    -- &  -7.48  & -7.03 & -6.68 & -6.23 & --  \\
      &     &    &       &  45   &  --  &    -- &  -7.56  & -7.12 & -6.63 & -6.22& --  \\
      &     &    &       &  40   &  -- &    -- &  -7.68  & -7.18 & -6.68 & -6.29 & -- \\
      &	    &    &       &  35   &  -- &    -- &  -7.56  & -7.09 & -6.76 & -6.45 & -- \\
      &     &    &       &  30   &  -- &  -7.98 & -7.45  & -7.19 & -6.92 & -6.60 & -- \\
\noalign{\smallskip}
      &     &    & 2.0   &  50   &  -- &  -7.79 & -7.25  & -6.88 & -6.46 & -6.01 & -- \\
      &     &    &       &  45   &  -- &  -7.93 & -7.35  & -6.91 & -6.47 & -5.97 & -- \\
      &     &    &       &  40   &  -- &  -8.16 & -7.47  & -7.01 & -6.48 & -6.05 & -- \\
      &	    &    &       &  35   &  -- &  -8.45 & -7.31  & -6.93 & -6.59 & -6.29 & -- \\
      &     &    &       &  30   &  -- &  -7.74 & -7.31  & -7.08 & -6.76 & -6.38 & -- \\
      &     &    &       &  27.5 &  -- &  -7.71 & -7.40  & -7.12 & -6.73 & -6.26 & -- \\          
      &     &    &       &  25   &  -- &  -7.76 & -7.42  & -7.04 & -6.48 & -6.01 & -- \\
      &	    &    &       &  22.5 &  -- &  -7.75 & -7.40  & -6.84 & -6.32 & -5.99 & -- \\
      &     &    &       &  20   &  -- &  -7.71 & -7.24  & -6.72 & -6.41 & -6.06 & -- \\
      &     &    &       &  17.5 &  -- &  -7.66 & -7.24  & -6.88 & -6.49 & -6.12 & -- \\          
      &     &    &       &  15   &  -- &  -7.88 & -7.42  & -6.98 & -6.62 & -6.15 & -- \\
      &	    &    &       &  12.5 &  -- &  -8.10 & -7.61  & -7.27 & -6.74 & -6.13 & -- \\
\noalign{\smallskip}
      &     &    & 1.3   &  22.5 &  -- &  -7.49 & -6.96  & -6.55 & -6.15 & -5.75 & -- \\
      &     &    &       &  20   &  -- &  -7.43 & -6.99  & -6.53 & -6.22 & -5.83 & -- \\
      &     &    &       &  17.5 &  -- &  -7.50 & -7.06  & -6.63 & -6.28 & -5.83 & -- \\          
      &     &    &       &  15   &  -- &  -7.53 & -7.22  & -6.85 & -6.39 & -5.79 & -- \\
      &	    &    &       &  12.5 &  -- &  -7.71 & -7.41  & -7.04 & -6.32 & -5.72 & -- \\
\noalign{\smallskip}
\hline
\noalign{\smallskip}
0.206 & 5.5 & 40 & 2.6   &  50   &  -- & -7.30 & -6.91 & -6.36 & -5.97 & -5.53 & -- \\
      &     &    &       &  45   &  -- & -7.30 & -7.12 & -6.41 & -5.95 & -5.45 & -- \\
      &     &    &       &  40   &  -- & -7.45 & -6.74 & -6.47 & -5.95 & -5.53 & -- \\
      &	    &    &       &  35   &  -- & -7.74 & -6.92 & -6.37 & -6.06 & -5.77 & -- \\
      &     &    &       &  30   &  -- & -7.10 & -6.80 & -6.58 & -6.25 & -5.90 & -- \\
\noalign{\smallskip}
      &     &    & 2.0   &  50   &  -- & -6.97 & -6.56 & -6.20 & -5.76 & -5.28 & -- \\
      &     &    &       &  45   &  -- & -7.02 & -6.65 & -6.22 & -5.73 & -5.24 & -- \\          
      &     &    &       &  40   &  -- & -7.10 & -6.73 & -6.26 & -5.75 & -5.35 & -- \\
      &	    &    &       &  35   &  -- & -7.33 & -6.70 & -6.27 & -5.90 & -5.60 & -- \\
      &     &    &       &  30   &  -- & -6.96 & -6.70 & -6.41 & -6.10 & -5.67 & -- \\
      &     &    &       &  27.5 &  -- & -7.04 & -6.78 & -6.48 & -6.01 & -5.56 & -- \\          
      &     &    &       &  25   &  -- & -7.09 & -6.79 & -6.38 & -5.75 & -5.34 & -- \\
      &	    &    &       &  22.5 &  -- & -7.07 & -6.62 & -6.12 & -5.66 & -5.33 & -- \\
      &     &    &       &  20   &  -- & -6.97 & -6.52 & -6.11 & -5.75 & -5.40 & -- \\
      &     &    &       &  17.5 &  -- & -6.88 & -6.59 & -6.17 & -5.86 & -5.43 & -- \\          
      &     &    &       &  15   &  -- & -7.03 & -6.78 & -6.35 & -5.93 & -5.43 & -- \\
      &	    &    &       &  12.5 &  -- & -7.35 & -6.96 & -6.70 & -6.09 & -5.31 & -- \\
\noalign{\smallskip}
      &     &    & 1.3   &  22.5 &  -- & -6.76 & -6.27 & -5.82 & -5.44 & -5.12 & -- \\
      &     &    &       &  20   &  -- & -6.61 & -6.28 & -5.88 & -5.52 & -5.18 & -- \\
      &     &    &       &  17.5 &  -- & -6.69 & -6.40 & -6.02 & -5.59 & -5.11 & -- \\          
      &     &    &       &  15   &  -- & -6.82 & -6.51 & -6.13 & -5.67 & -5.03 & -- \\
      &	    &    &       &  12.5 &  -- & -7.06 & -6.78 & -6.26 & -5.65 & -4.92 & -- \\
\noalign{\smallskip}
\hline
\end{tabular}
\end{table*}

\addtocounter{table}{-1}%

\begin{table*}
\caption[]{{\bf Continued.} Predicted mass-loss rates for different metallicities.}
\label{t_massloss2}
\begin{tabular}{ccccrccccccc}
\hline
\noalign{\smallskip}
\multicolumn{5}{c}{} & \multicolumn{5}{c}{log $\dot{M}(\msunyr)$} \\
\noalign{\smallskip}
$\Gamma_{\rm e}$ & log$L_*$ & $M_*$ & $\ratio$ & $\teff$ & 1/100 & 1/30 & 1/10 & 1/3 & 1 & 3 & 10 \\
\noalign{\smallskip}
         & ($\lsun$) & ($\msun$) & & (kK) & $Z/\zsun$ &$Z/\zsun$ & $Z/\zsun$ & $Z/\zsun$ & $Z/\zsun$ & $Z/\zsun$ & $Z/\zsun$ \\
\hline
0.434 & 6.0 & 60 & 2.6   &  50   & -6.81 & -6.31 & -5.84 & -5.46 & -5.07 & -4.57 & -4.31   \\
      &     &    &       &  45   & -6.80 & -6.59 & -5.87 & -5.45 & -4.99 & -4.55 & -4.31 \\
      &     &    &       &  40   & -6.86 & -6.16 & -5.95 & -5.41 & -4.97 & -4.59 & -4.42  \\
      &	    &    &       &  35   & -7.16 & -6.27 & -5.95 & -5.47 & -5.05 & -4.78 & -4.60 \\
      &     &    &       &  30   & -6.78 & -6.21 & -5.90 & -5.57 & -5.29 & -4.94 & -4.52  \\
\noalign{\smallskip}
      &     &    & 2.0   &  50   & -6.42 & -6.17 & -5.67 & -5.25 & -4.86 & -4.42 & -4.23 \\
      &     &    &       &  45   & -6.47 & -6.35 & -5.69 & -5.22 & -4.76 & -4.42 & -4.24 \\          
      &     &    &       &  40   & -6.58 & -5.98 & -5.73 & -5.23 & -4.76 & -4.47 & -4.32  \\
      &	    &    &       &  35   & -6.78 & -6.11 & -5.74 & -5.28 & -4.88 & -4.65 & -4.47  \\
      &     &    &       &  30   & -6.47 & -6.07 & -5.80 & -5.44 & -5.14 & -4.82 & -4.38 \\
      &     &    &       &  27.5 & -6.50 & -6.16 & -5.99 & -5.51 & -5.19 & -4.68 & -4.23  \\          
      &     &    &       &  25   & -6.60 & -6.24 & -5.92 & -5.38 & -4.95 & -4.44 & -4.11 \\
      &	    &    &       &  22.5 & -6.52 & -6.11 & -5.63 & -5.13 & -4.78 & -4.45 & -4.17 \\
      &     &    &       &  20   & -6.33 & -5.93 & -5.59 & -5.19 & -4.83 & -4.54 & -4.26 \\
      &     &    &       &  17.5 & -6.36 & -6.01 & -5.74 & -5.33 & -4.90 & -4.48 & -4.11 \\          
      &     &    &       &  15   & -6.54 & -6.17 & -5.90 & -5.42 & -4.85 & -4.25 & -3.94 \\
      &	    &    &       &  12.5 & -6.71 & -6.35 & -5.99 & -5.48 & -4.51 & -4.19 & -3.99 \\
\noalign{\smallskip}
      &     &    & 1.3   &  22.5 & -6.24 & -5.77 & -5.36 & -4.91 & -4.55 & -4.29 & -4.10 \\
      &     &    &       &  20   & -6.06 & -5.70 & -5.37 & -5.00 & -4.63 & -4.38 & -4.12 \\
      &     &    &       &  17.5 & -6.09 & -5.80 & -5.52 & -5.09 & -4.59 & -4.19 & -3.97 \\          
      &     &    &       &  15   & -6.29 & -5.98 & -5.65 & -5.07 & -4.28 & -4.06 & -3.91 \\
      &	    &    &       &  12.5 & -6.49 & -6.13 & -5.75 & -4.80 & -4.30 & -4.10 &  -3.95 \\
\noalign{\smallskip}
\hline
\end{tabular}
\end{table*}

\section{The predicted mass-loss rates and bi-stability jumps}
\label{s_massloss}

\begin{figure*}
\centerline{\psfig{file=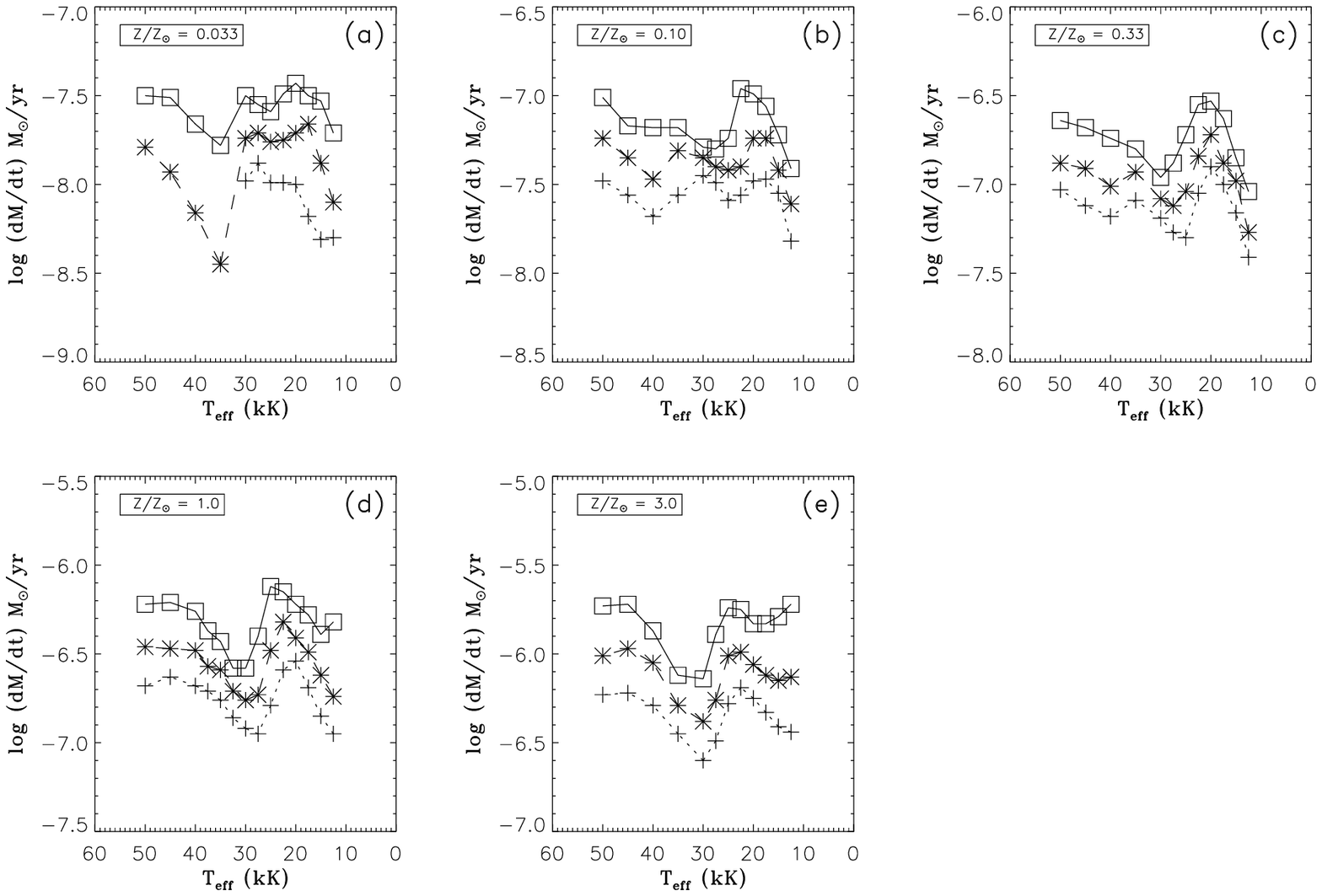, width = 18.0cm}}
\centerline{\psfig{file=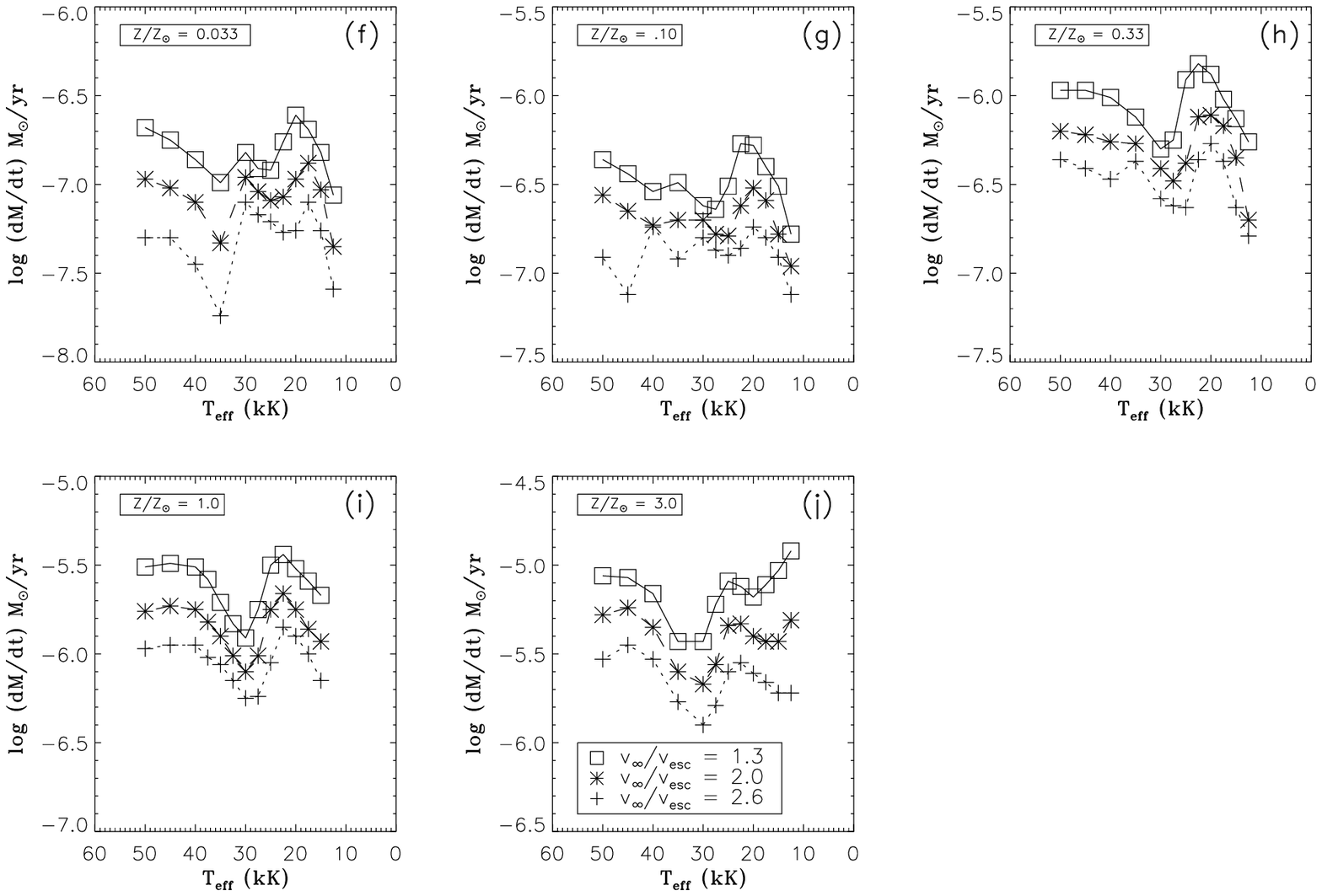, width = 18.0cm}}
\caption{The calculated mass-loss rates $\dot{M}$ as a 
function of $\teff$ for five metallicities in the range $Z/\zsun$ = 1/30 - 3.
The metal content is indicated in the legend at the upper part of each panel.
Upper five panels (a)-(e): $\Gamma_{\rm e}$ = 0.130 (log L/$\lsun$ = 5.0).
Lower five panels (f)-(j): $\Gamma_{\rm e}$ = 0.206 (log L/$\lsun$ = 5.5).
The values for ($\ratio$) are indicated in the legend at the lower part of the last panel (j).} 
\label{f_mdotL5}
\end{figure*}

\addtocounter{figure}{-1}%

\begin{figure*}
\centerline{\psfig{file=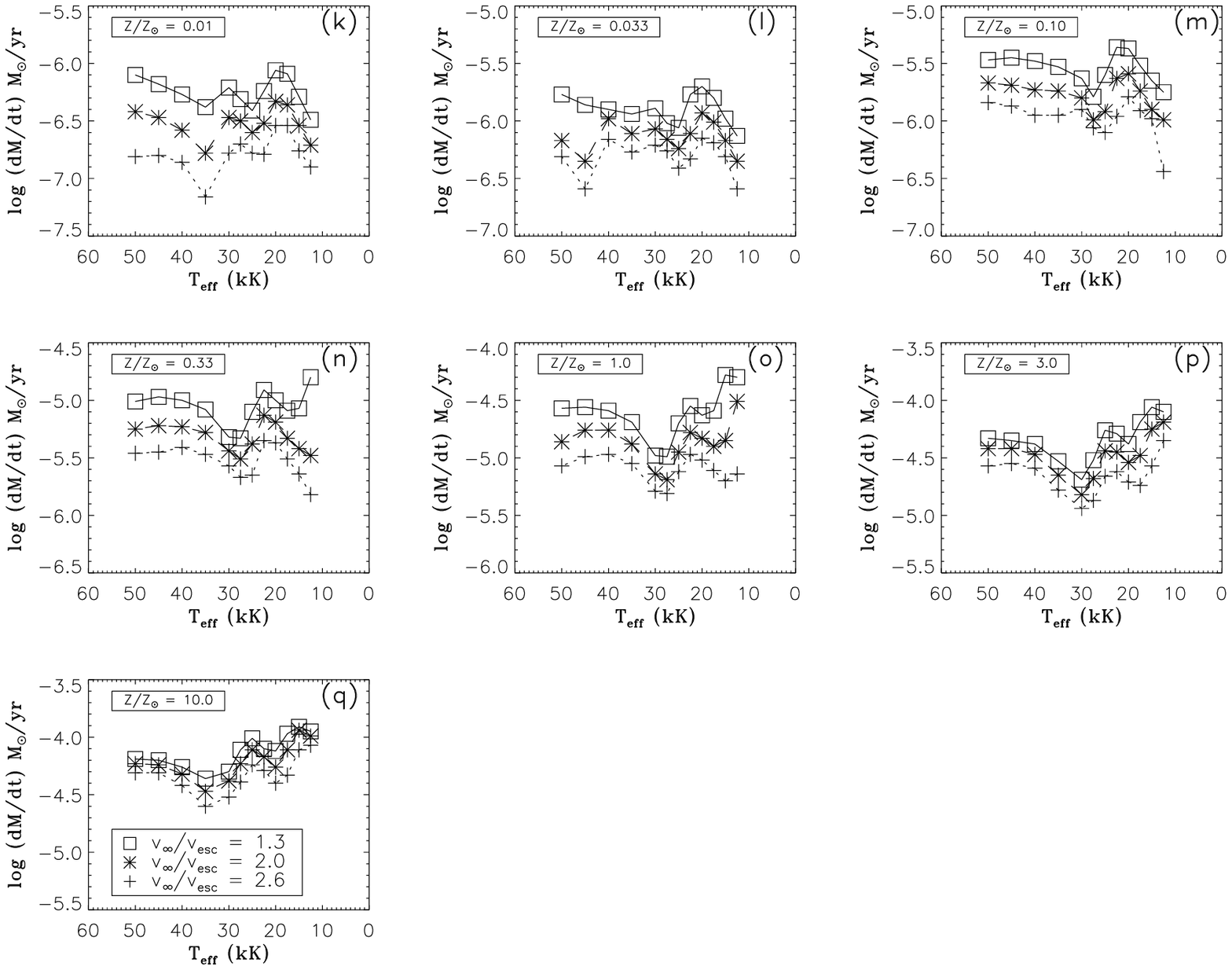, width = 18.0cm}}
\caption{{\bf Continued.} Series of $\dot{M}(Z)$ calculations with $\Gamma_{\rm e}$ = 0.434 (log L/$\lsun$ = 6.0).
The calculated mass loss as a function of $\teff$ for seven metallicities in the range $Z/\zsun$ = 1/100 - 10.
The metal abundance is indicated in the legend at the upper part of each panel (k-q).
The values for ($\ratio$) are indicated in the legend of the last panel (q).} 
\label{f_mdotL6}
\end{figure*}

The calculated mass-loss rates are shown in 
the different panels of Fig.~\ref{f_mdotL5} and most results 
are also given in Table~\ref{t_massloss}.
They show bi-stability jumps superimposed on 
an overall behaviour where $\dot{M}$ decreases for decreasing  
$\teff$. The reason for this $\dot{M}$ decrease is that the maximum of 
the flux distribution gradually shifts to longer wavelengths. 
Since there are significantly less 
lines at roughly $\lambda \ga$ 1800 \AA\ than at shorter wavelength, 
the line acceleration becomes less effective at lower 
$T_{\rm eff}$, and thus the mass loss decreases. 

However, most of the panels of Fig.~\ref{f_mdotL5} show bi-stability 
jumps, where the mass loss drastically increases.
Before we can investigate the overall dependence of metallicity on mass loss,
we need to describe the positions of these bi-stability jumps in $\teff$.

\subsection{The bi-stability jump at $\teff \simeq 25~000$ K}

All panels show a bi-stability jump around $\teff$ $\simeq$ 25~000 K. 
Here, Fe {\sc iv} recombines 
to Fe {\sc iii} and as the latter ion is a more efficient line driver 
than the first, the acceleration in the lower part of the 
wind increases. This results in an upward jump in $\dot{M}$ of about a 
factor of five and subsequently a drop in $\vinf$ of about a factor 
0.5 (Vink et al. 1999).

Since we know from both theory and observations 
that the Galactic ratio $\ratio$ jumps from $\sim$ 2.6 at the hot 
side of the jump to $\sim$ 1.3 at the cool side of the jump, 
we can estimate the size of the jump in mass loss
for the different metallicities by assuming a similar jump in the 
ratio $\ratio$ of about a factor of two. The size of 
the jump is defined as the difference between the minimum $\dot{M}$ 
at the hot side of the jump (where $\ratio$ = 2.6) 
and the maximum $\dot{M}$ at the cool side (where $\ratio$ = 1.3).
The size of the predicted jump in $\dot{M}$ (i.e. $\Delta$ log $\dot{M}$) is 
indicated in the last column of Table~\ref{t_jumps}. For most models 
$\Delta\dot{M}$ is about a factor of five to seven. There is no clear 
trend with metallicity.

The position of the jump for different $Z$ shifts somewhat in 
$\teff$, since the ionization equilibrium does not only 
depend on temperature, but also on density and 
therefore on mass loss and thus on metallicity as well. 
To handle the influence of the metallicity on the position
of the bi-stability jump in $\teff$, we compare the 
characteristics of the wind models around the bi-stability jump. 
We will discuss this behaviour for the case of the highest wind 
densities ($\Gamma_{\rm e}$ = 0.434), as for these models, the 
statistics in the Monte-Carlo code are
the best (see Sect.~\ref{s_method}). Nevertheless, the uniformity is checked for 
the other series of $\Gamma_{\rm e}$ also.

\begin{table}
\caption[]{The size of the bi-stability jump around 25~000 K for different $Z$.}
\label{t_jumps}
\begin{tabular}{ccclc}
\hline
\noalign{\smallskip}
$\Gamma_{\rm e}$ & log$L_*$ & $M_*$ & ($Z/\zsun$) & $\Delta$ (log  $\dot{M}$) \\
\noalign{\smallskip}
         & ($\lsun$) & ($\msun$) &    &  \\
\hline
0.130 & 5.0 & 20 & 1/30 &   -   \\
      &     &    & 1/10  &  0.75 \\
      &     &    & 1/3  &  0.77 \\
      &	    &    & 1     &  0.83 \\
      &     &    & 3     &  0.86 \\
0.206 & 5.5 & 40 & 1/30  &  0.66 \\
      &     &    & 1/10   &  0.63 \\
      &     &    & 1/3  &  0.81 \\
      &	    &    & 1     &  0.81 \\
      &     &    & 3     &  0.81 \\	
0.434 & 6.0 & 60 & 1/100  &  0.72 \\
      &     &    & 1/30  &  0.71 \\
      &     &    & 1/10   &  0.74 \\
      &     &    & 1/3    &  0.76 \\
      &     &    & 1     &  0.76 \\
      &     &    & 3     &  0.68 \\
      &     &    & 10    &  0.43 \\
\noalign{\smallskip}
\hline
\end{tabular}
\end{table}

\begin{figure}
\centerline{\psfig{file=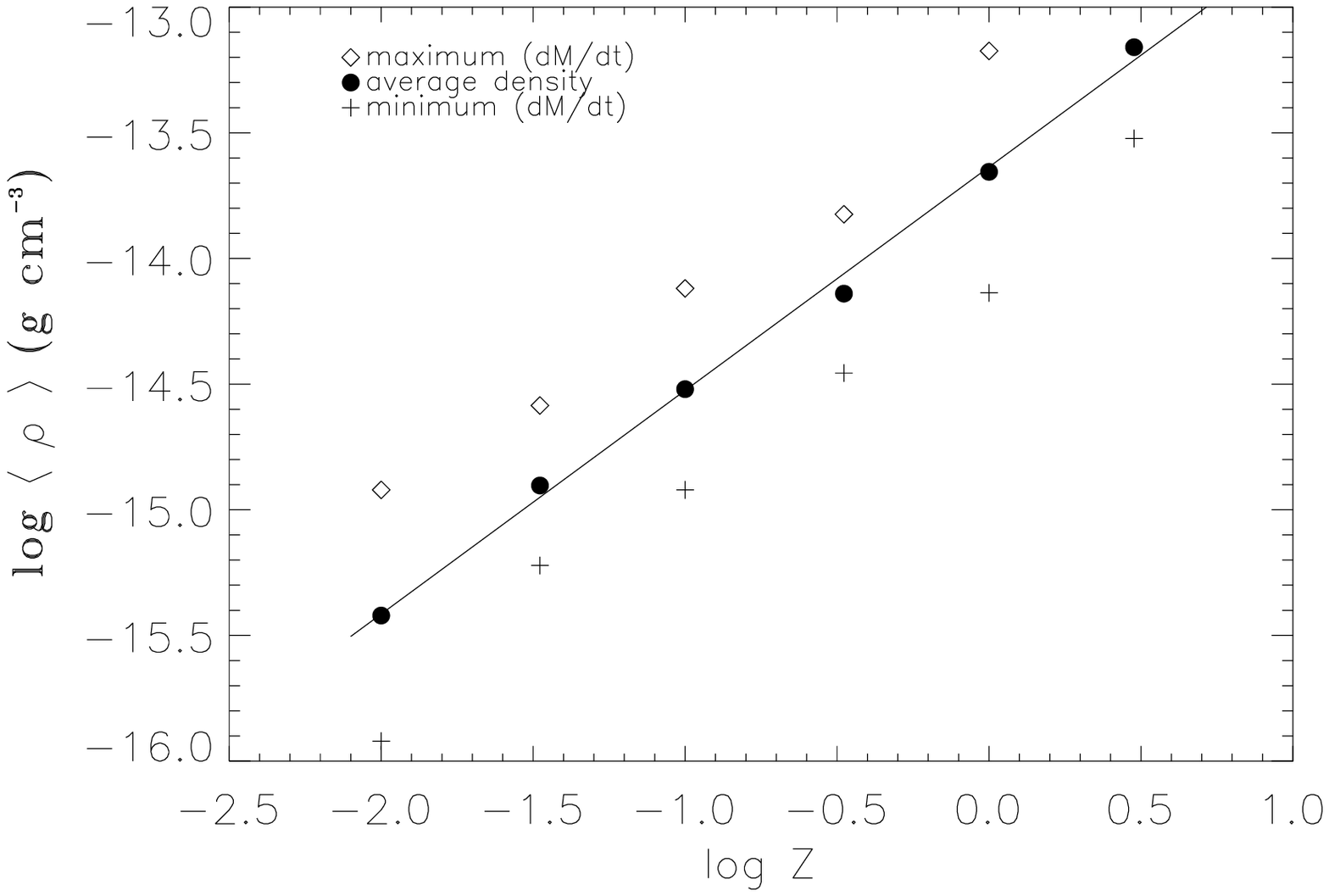, width = 9 cm}}
\caption{Characteristic density $\langle\rho\rangle$ at the bi-stability jump around 25~000 K as a function of $Z$.
An explanation for the different symbols is given in the legend. The solid
line indicates the best linear fit through the average jumps parameters for log $\langle\rho\rangle$.}
\label{f_logZ}
\end{figure}

\begin{figure}
\centerline{\psfig{file=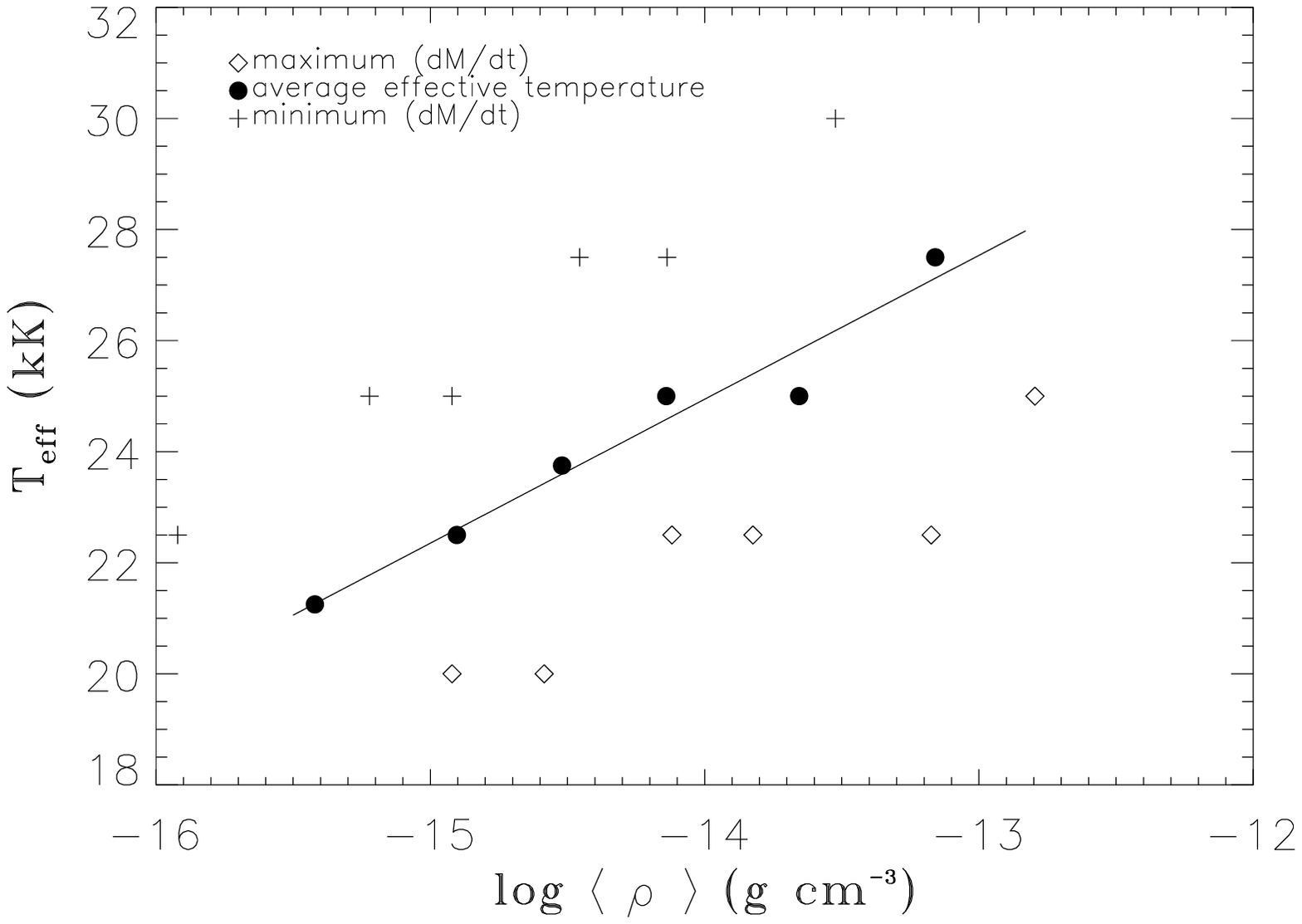, width = 9 cm}}
\caption{Characteristic density log $\langle\rho\rangle$ and $\Teff$ of the bi-stability jump around $\teff$ = 25~000 K.
An explanation for the different symbols is given in the legend. The solid
line represents the best linear fit through the average jump parameters log $\langle\rho\rangle$ and $\teff$.}
\label{f_jump}
\end{figure}

As in Vink et al. (2000), $\langle\rho\rangle$ is defined as the characteristic wind 
density at 50 \% of the terminal velocity of the wind. 
For a standard velocity law with $\beta=1$, this characteristic wind density is given by

\begin{equation}
\langle\rho\rangle~=~\frac{\dot{M}}{8 \pi R_*^2 \vinf}
\label{eq_cdens}
\end{equation}
Figure~\ref{f_logZ} shows the behaviour of the characteristic 
density as a function of $Z$. This is done for both the minimum 
$\dot{M}$ (at the hot side of the jump) and the maximum
$\dot{M}$ (at the cool side of the jump). The characteristic densities 
for the cool side of the jump are indicated with ``diamond'' signs and with ``plus'' signs 
for the hot side. The ``filled circles'' represent the logarithmic average 
values of $\langle\rho\rangle$ for the ``jump'' model for each metallicity. 
The ``jump'' model is a hypothetical model between the two models 
where $\dot{M}$ is maximal and minimal. 
As expected, log $\langle\rho\rangle$ increases as the metallicity increases.
Because the log of the average density at the jump
shows a linear dependence on log ($Z/\zsun$), a linear fit 
is plotted. This is the solid line in Fig.~\ref{f_logZ}.
The relation is given by:

\begin{eqnarray}
{\rm log}< \rho >~& = &~- 13.636~(\pm~0.029)~\nonumber\\
                  & & ~+~0.889~(\pm~0.026)~{\rm log} (Z/\zsun)
\label{eq_logZ}
\end{eqnarray}
Figure~\ref{f_jump} shows the effective temperature of the bi-stability jump
as a function of $\langle\rho\rangle$. Again this is done for both
the cool and hot side of the jump and for the average.
The solid line indicates the best linear fit through these averages. 
The relation between the jump temperature (in kK) and log $\langle\rho\rangle$ is given by:

\begin{equation}
T_{\rm eff}^{\rm jump}~=~61.2~(\pm~4.0)~+~2.59~(\pm~0.28)~{\rm log} < \rho > 
\label{eq_jump}
\end{equation}
It is now possible to estimate
$\langle\rho\rangle$ for any $Z$ using Eq.~(\ref{eq_logZ}) and subsequently to
predict the position of the jump in $\teff$ using Eq.~(\ref{eq_jump}).

\subsection{Additional bi-stability jumps around 15~000 and 35~000 K}

In many of the panels in Fig.~\ref{f_mdotL5} one can see 
more than just one bi-stability jump. In cases for high mass loss at
relatively high $Z$, an additional jump is visible at $\teff \simeq$ 
15~000 K (see e.g. panel (o) in Fig.~\ref{f_mdotL5}). 
Leitherer et al. (1989) calculated atmospheric models for
Luminous Blue Variables (LBVs) and found a recombination of iron group 
elements from doubly to singly ionised stages, which may explain
mass-loss variability when LBVs change from minimum to maximum 
visual brightness phase (de Koter et al. 1996). 
Vink et al. (2000) also found this jump around 15~000 K and attributed it 
to a recombination of Fe {\sc iii} to Fe {\sc ii}. Possibly this jump is related to
the drop in the ratio $\ratio$ from 1.3 to about 0.7 around spectral type A0 
as identified by Lamers et al. (1995) on the basis of observed values for $\vinf$.

For the lower mass-loss rates at relatively low metallicity, at about
$\teff \simeq$ 35~000 K, another
drastic increase in $\dot{M}$ occurs (e.g. panel (f) with 
$Z/\zsun$ = 1/30 in Fig.~\ref{f_mdotL5}).
The origin of this 35~000 K jump, which appears only at low $Z$, will 
be discussed in Sect.~\ref{s_extrajump}. 

In order to express the mass-loss behaviour as a function of metal
content, it is obvious that
all these jumps need to be accounted for. Since these additional
jumps are only present in a few cases, the relationships can only be  
given as rough estimates. For the jump at $\teff \simeq$ 15~000 K:

\begin{equation}
T^{\rm jump}_{\sim15{\rm kK}}~=~43~+~1.9~{\rm log} < \rho > 
\label{eq_jump2}
\end{equation}
For the jump at $\teff \simeq$ 35~000K:

\begin{equation}
T^{\rm jump}_{\sim35{\rm kK}}~=~192~+~10.4~{\rm log} < \rho > 
\label{eq_jump3}
\end{equation}
In both cases the jump temperature is in units of kK.
It is again possible to estimate 
log $\langle\rho\rangle$ using Eq.~(\ref{eq_logZ}) and then to roughly
predict the positions of these additional bi-stability jumps in 
effective temperature using Eqs.~(\ref{eq_jump2}) 
and~(\ref{eq_jump3}). Later on these will be referred to 
when the complete mass-loss recipe is presented (Sect.~\ref{s_recipe}).

\subsection{The origin of the (low $Z$) jump at $\teff \simeq$ 35~000 K}
\label{s_extrajump}

Intuitively, one might attribute the jump at $\sim$ 35~000 K in models
of low metal abundance (say $Z/\zsun \le 1/30$) to the recombination of 
Fe {\sc v} to Fe {\sc iv}. This in analogue to the jump at $\sim$
25~000 K, due to the recombination of Fe {\sc iv} to Fe {\sc iii}.
However, in the next section we will show 
that this is not the case, since at lower $Z$ the relative contribution of Fe 
vs. CNO in the line acceleration decreases (see also Puls et al. 2000).

Instead, the low $Z$ jump at $\teff \simeq 35~000$ K turns out 
to be caused by a recombination from carbon {\sc iv} to carbon {\sc iii}
(see Vink 2000). To summarise the physical origin of the jump: 
C {\sc iii} has more lines in the crucial part 
of the spectrum than C {\sc iv}, therefore C {\sc iii} is a more efficient driving 
ion causing the increase in mass loss at the bi-stability jump
around 35~000 K at low $Z$. Whether this is also 
accompanied by a change in terminal velocity is an open question 
that may be answered if $\vinf$ 
determinations at very low $Z$ become available.

\section{The relative importance of Fe and CNO elements in the line acceleration at low $Z$}
\label{s_relabund}

\subsection{The character of the line driving at different $Z$}

\begin{figure*}
 \centerline{\psfig{file=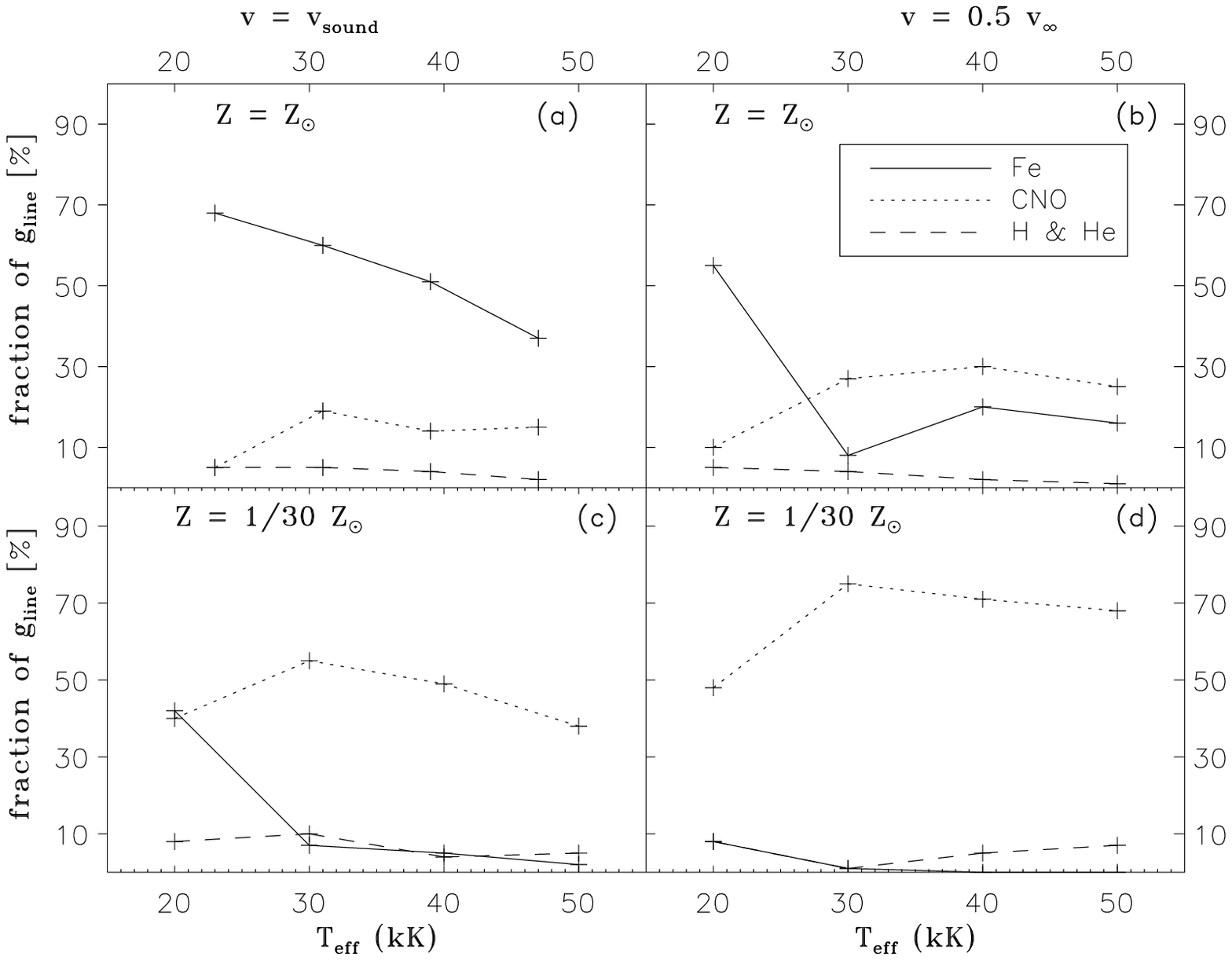, width = 16 cm}}
 \caption{The relative contribution to the line acceleration
           for models with ($\ratio$) = 2.0, log $L_{*}/\lsun$ = 5.5 and 
           $M_* = 40 \msun$.
           The solid lines show the contribution of Fe lines.
           The dotted line is the contribution by  
           C, N and O. The dashed line 
           shows the contribution by H and He lines. 
           {\bf (a)} and {\bf (b)}~give the contribution for solar $Z$ at 
           resp.  $v~=~v_{\rm sound}$ and at $v~=~0.5 \vinf$.
           {\bf (c)} and {\bf (d)}~give the contribution for  ($Z/\zsun$) = 1/30 
           at resp.  $v~=~v_{\rm sound}$ and at $v~=~0.5 \vinf$.}
\label{f_Zrel}
\end{figure*}

Vink et al. (1999) have shown that for Galactic wind models 
around 25~000 K the elements C, N and O are important line drivers 
in the supersonic part of the wind, whereas the subsonic part of the 
wind is dominated by the line acceleration
due to Fe. As the mass-loss rate is determined by the radiative acceleration
below the sonic point, and the terminal velocity is determined
by the acceleration in the supersonic part, these
results imply that for Galactic wind models $\dot{M}$ is essentially set 
by Fe lines, whereas $\vinf$ is determined by the lighter elements, i.e. mainly by CNO.

To study the origin of the additional (low $Z$) jump around 
35~000 K, it becomes necessary to investigate the relative importance of the 
species at low metallicity. To this end, additional 
Monte Carlo calculations were performed.
One simulation was performed with a line list containing only Fe lines. A second calculation
was done with a list of lines of CNO only, and finally a third simulation was performed with
the lines of H and He.
Figure~\ref{f_Zrel} shows the relative importance for the line acceleration of these elements 
as a function of effective temperature for different parts of the wind, i.e. at $v~=~v_{\rm sound}$ 
and at $v~=~0.5 \vinf$. Panel (a) and (b)
indicate the fractions in the acceleration at 
solar metallicity. Panel (c) and (d) present the same, but for the low metallicity models, i.e.  
$Z/\zsun = 1/30$. Note that for the solar metallicity models in the supersonic region (panel b)
the elements of Si, Cl, P and S are additional line drivers (see Vink et al. 1999).

Figure~\ref{f_Zrel}(a) shows that at solar $Z$, Fe dominates the line 
acceleration around the sonic point, where the mass-loss rate is fixed.
However, this relative importance of iron decreases for increasing $\teff$.
Figure~\ref{f_Zrel}(c) shows that at the low metallicity, CNO already dominate 
the acceleration in the region around the sonic point. This implies that 
at low $Z$, CNO determine both the terminal velocity by dominating the supersonic line acceleration 
in Fig.~\ref{f_Zrel}(d), as well as the mass loss by dominating the line 
acceleration around  $v~=~v_{\rm sound}$. The only exception occurs at low 
effective temperature ($\teff =$ 20~000 K), where Fe still plays an important role in
setting the mass loss.

These considerations thus explain why the high $\teff$ jump at low $Z$ is not 
caused by a recombination effect of iron, instead it turns out to 
be caused by a recombination
of a CNO element, in this case C {\sc iv} to C {\sc iii} (Vink 2000).

\subsection{Observed abundance variations at different $Z$}
\label{s_obsabund}

Now we will make a distinction between 
the metal abundance $Z$ derived on the basis of stellar iron and nebular 
oxygen lines. The reason for this distinction is that observations to 
study the chemical evolution of galaxies have shown that the 
ratio of Fe/O varies with metallicity.

Determinations of heavy-element abundances for metal poor blue compact 
galaxies (Izotov \& Thuan 1999) as well as observations of 
Galactic halo stars (Pagel \& Tautvaisiene 1995 and references therein)
show a significant overabundance of O/Fe of about 0.4 dex with respect to the Sun. 

These observed differential abundance variations between oxygen
and iron could significantly alter our mass-loss predictions 
if $\dot{M}$ were set by Fe over the full range in $Z$.
However, we have shown that at low $Z$, the mass loss
is mainly determined by CNO instead of by Fe. Since the observed
metallicity is mostly determined from nebular oxygen lines rather than 
from iron lines, this implies that our mass loss recipe will still yield 
the proper mass-loss rates. Only in those cases where the 
observed metallicity were determined from stellar iron lines instead 
of from nebular oxygen lines, one would need to transform the observed 
iron abundance ($Z^{\rm obs}_{\rm Fe}$) to our adopted metallicity 
($Z^{\rm theory}$). 
This can easily be done according to the scaling relations 
given in Table~\ref{t_obsabund}. The first column of this table indicates the 
metallicity that has been adopted in the wind models. The 
second column shows for each $Z$ which elements dominate the line driving
around the sonic point, where the mass loss is set. The third column 
represents the observed abundance variation between oxygen and iron
compared to the sun. For relatively high metallicity ($Z/\zsun \ga$ 1/10), there
is hardly any observed difference between the oxygen and iron abundances.
As said, for very low metallicity ($Z/\zsun \la$ 1/30), this observed difference is 
about 0.4 dex. 
Because at low $Z$ mass loss is mainly set by CNO,
the observed oxygen abundances are the same as the adopted $Z$ in
the wind models (column 4), whereas in case iron lines were to be  
analysed, one should convert the iron abundance to our adopted $Z^{\rm theory}$, by 
comparing column 5 and column 1.

\begin{table}
\caption[]{Conversion table for the observed differential abundance variations 
           between oxygen and iron. }
\label{t_obsabund}
\begin{tabular}{ccccc}
\hline
\noalign{\smallskip}
$Z^{\rm theory}$ & Dominant elements & $\frac{[O/Fe]}{[O/Fe]_{\odot}}$ & $Z^{\rm obs}_{\rm Oxygen}$ & $Z^{\rm obs}_{\rm Fe}$ \\
\noalign{\smallskip}
($\zsun$) & that set $\dot{M}$ & &($\zsun$) & ($\zsun$) \\
\hline
1      & Fe  &   0     &  1    & 1    \\
1/3    & Fe  &   0     &  1    & 1    \\
1/10   & Fe  &   0     &  1    & 1    \\
\noalign{\smallskip}
\hline
\noalign{\smallskip}
1/30  & CNO & + 0.4 dex & 1/30  & 1/75  \\
1/100 & CNO & + 0.4 dex & 1/100 & 1/250 \\
\noalign{\smallskip}
\hline
\end{tabular}
\end{table}

\section{The global metallicity dependence}
\label{s_metal}

\begin{figure}
\centerline{\psfig{file=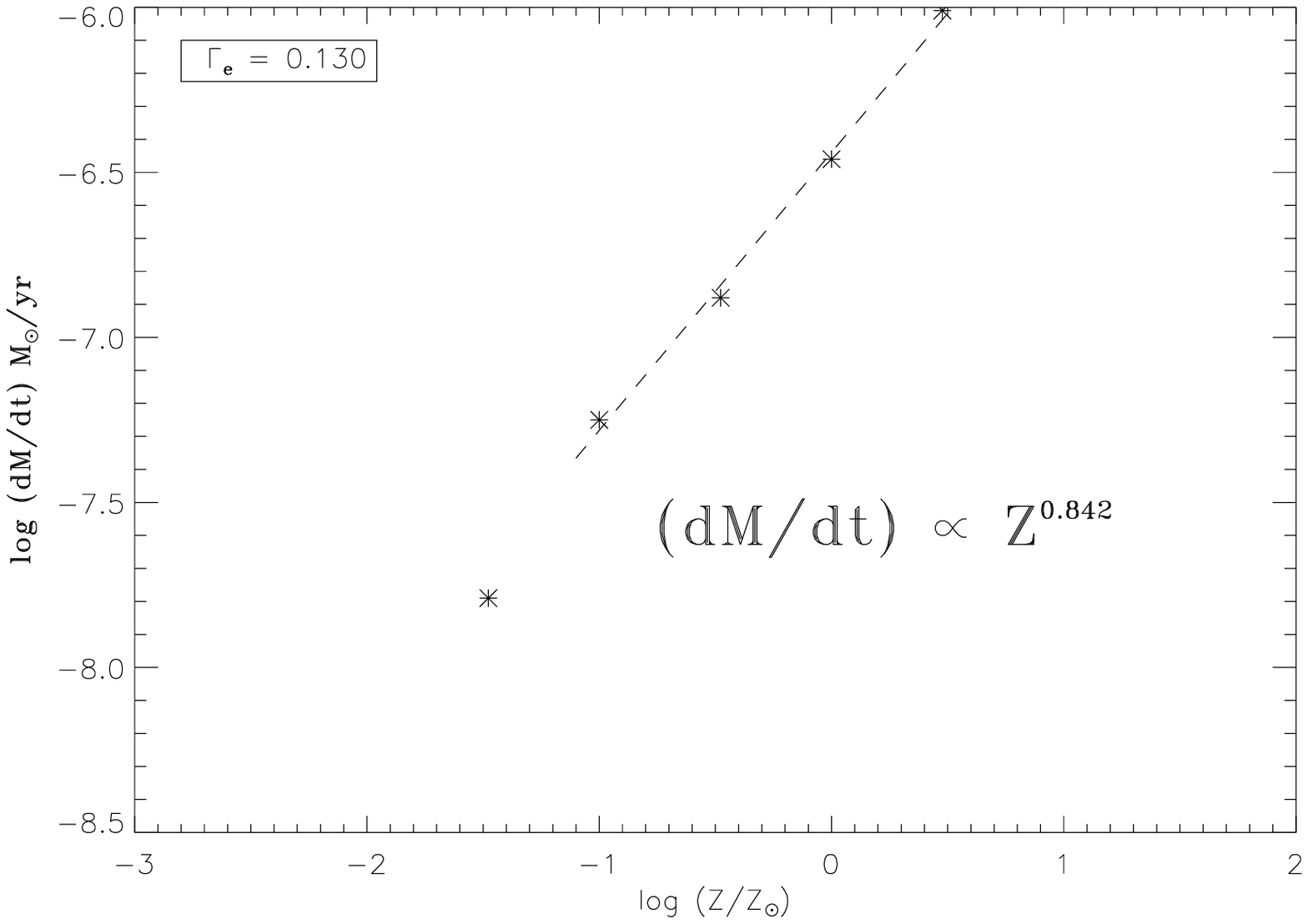, width = 9 cm}}
\centerline{\psfig{file=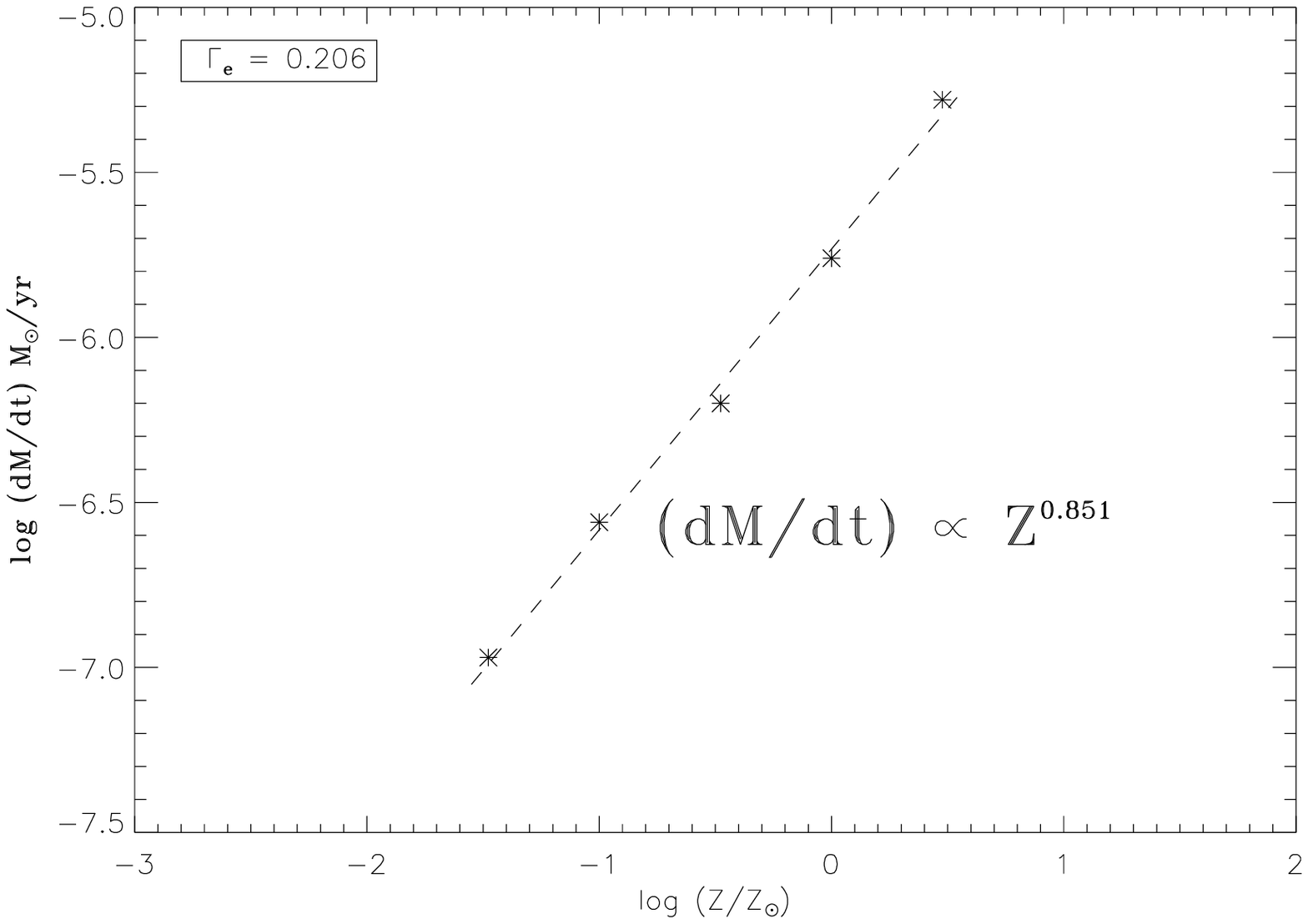, width = 9 cm}}
\centerline{\psfig{file=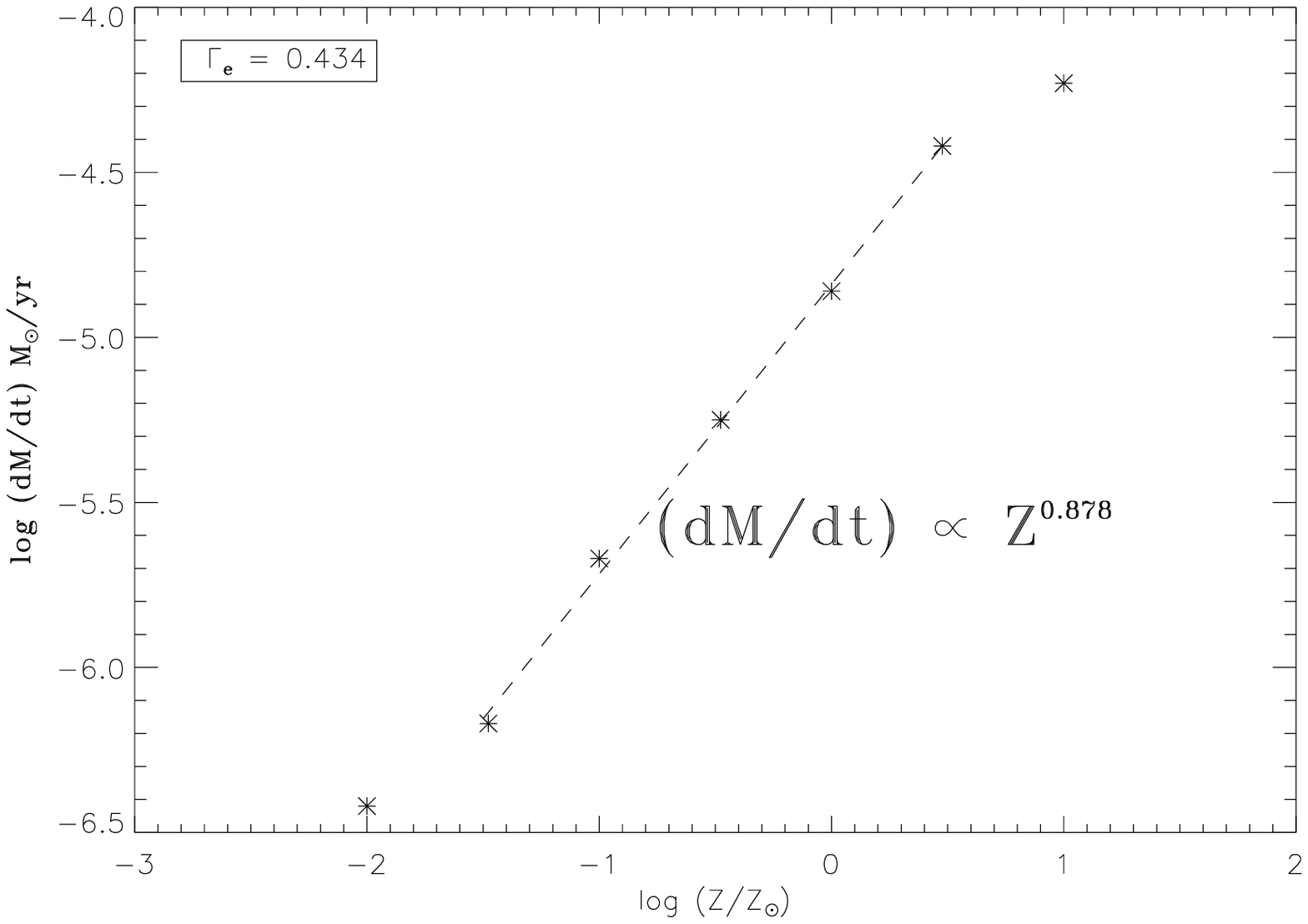, width = 9 cm}}
\caption{The $\dot{M}(Z)$ dependence for three cases of $\Gamma_{\rm e}$.
In all three panels, the dashed lines indicate the best linear  
fit through the models at different $Z$.
Note that at $\Gamma_{\rm e} = 0.130$ the lowest $Z$ model is 
not included in the fit, due to the presence of a bi-stability jump.
All models have $\teff$ = 50~000 K and {\bf constant} ($\ratio$) = 2.0.
The values of $\Gamma_{\rm e}$ are indicated in the legends.}
\label{f_metal}
\end{figure}

Now we can determine the global $\dot{M}(Z)$ dependence over
a wide range in metallicity. This $\dot{M}(Z)$ will be determined 
for the three $\Gamma_{\rm e}$ values separately. 
If the dependencies were identical for different $\Gamma_{\rm e}$,
then we might simply add the metallicity dependence to the mass-loss recipe 
that was derived by Vink et al. (2000) for Galactic stars.

Figure~\ref{f_metal} shows the $\dot{M}(Z)$ behaviour for the 
three cases of $\Gamma_{\rm e}$. To avoid complications
due to the presence of the bi-stability jumps, we use models 
where $\teff$ is above all of the identified jumps, i.e. at $\teff$ = 50~000 K. 
In the case where $\Gamma_{\rm e}$ = 0.130, the linear fit is taken in the 
metallicity range $Z/\zsun =$ 1/10 - 3, because the model at 
$Z/\zsun$ = 1/30 is influenced by the low $Z$
bi-stability jump. This is why we have excluded this from the fit.
The best linear fit is thus given by

\begin{eqnarray}
\label{eq_m0130}
{\rm log}~\dot{M}& = &~-6.439~(\pm 0.024) \nonumber\\
                 & & ~+~0.842~(\pm 0.039)~{\rm log}(Z/\zsun) \nonumber\\
                 \nonumber\\
                 & &~{\rm for}~\Gamma_{\rm e}~=~0.130
\end{eqnarray}
In case $\Gamma_{\rm e}$ = 0.206, the models at $\teff$ = 50~000 K are not influenced 
by the low $Z$ jump and a linear fit is taken over the 
full metallicity range of $Z/\zsun =$ 1/30 - 3. 
The best fit is given by

\begin{eqnarray}
\label{eq_m0206}
{\rm log}~\dot{M} & = &~-5.732~(\pm 0.028) \nonumber \\
                  & & ~+~0.851~(\pm 0.033)~{\rm log}(Z/\zsun) \nonumber\\
                 \nonumber\\
                 & &~{\rm for}~\Gamma_{\rm e}~=~0.206
\end{eqnarray}
Finally, in the case $\Gamma_{\rm e}$ = 0.434, the $\dot{M}(Z)$ dependence 
is studied over an even wider metallicity range: $Z/\zsun$ is 1/100 - 10.
For this relatively high value of $\Gamma_{\rm e}$ it is 
computationally easier to calculate mass loss at the extremely low value 
$Z/\zsun = 1/100$. The mass-loss rate at extremely high 
metallicity ($Z/\zsun$ = 10) is determined for a somewhat different 
abundance ratio than the standard one that was used throughout 
the paper given by Eq.~(\ref{eq_he}). The helium 
abundance is now kept constant (at $Y$ = 0.42, see Table~\ref{t_abund}) 
increasing the metal fraction from three to ten times solar. It was checked 
whether the results are dependent on this choice of Y, but this turned out 
not to be the case.

One may expect the $\dot{M}(Z)$ relation to flatten at some 
high $Z$ value due to saturation of iron lines (see Sect.~\ref{s_context}). 
The lowest panel in Fig.~\ref{f_metal} shows that 
this is indeed the case. However, this only happens above $Z/\zsun$ = 3. 
It implies that over the range from about $Z/\zsun =$~1/30 - 3, the wind 
momentum behaves as a {\it constant} function of metallicity, 
i.e. mass loss vs. $Z$ behaves as a power-law.
The linear fit for the highest value of $\Gamma_{\rm e}$ is determined from the 
range $Z/\zsun =$~1/30 - 3. 
The best fit is given by

\begin{eqnarray}
\label{eq_m0434}
{\rm log}~\dot{M} & = &~-4.84~(\pm 0.020) \nonumber\\
                  & & ~+~0.878~(\pm 0.023)~{\rm log}(Z/\zsun) \nonumber\\
                 \nonumber\\
                 & &~{\rm for}~\Gamma_{\rm e}~=~0.434
\end{eqnarray}
Combining Eqs.~(\ref{eq_m0130}), ~(\ref{eq_m0206}) and ~(\ref{eq_m0434})
for the three different values of $\Gamma_{\rm e}$, we find
that over the metallicity range from $1/30 \le Z/\zsun \le 3$ 
there is a constant power law for constant $\ratio = 2.0$ and $\teff$ = 50~000 K with 
$\dot{M} \propto Z^{0.86}$. 

We have done similar analyses for the other effective temperatures in
our model grid, some of these were affected by a bi-stability jump, but
on average, these jumps cancelled out. The average power-law index 
factor $m$ (Eq.~\ref{eq_mdot}) was found to be $m$ = 0.85 $\pm$ 0.10 for constant $\ratio$. 
As was shown in Vink et al. (2000) $\dot{M}$ depends on $\ratio$ as a power law: $\dot{M} \propto \vinf^p$, 
with $p = -1.226 \pm 0.037$ for stars with $\teff \ga 25~000$ K, and $p = -1.601 \pm 0.055$ for 
the B supergiants with $\teff \la 25~000$ K. Therefore mass loss can be represented by

\begin{eqnarray}
\dot{M}~\propto~Z^m~\vinf^p & \propto~& Z^{0.85}~\vinf^p\nonumber\\
                 \nonumber\\
       & &~{\rm for}~1/30~\le~Z/\zsun~\le~3
\label{eq_mdot2}
\end{eqnarray}
Because $\vinf$ also depends on the metal content $Z$, where Leitherer et al. (1992) 
have assumed $\vinf$ to behave as a power-law with $\vinf \propto Z^q$ and 
derived this value to be $q = 0.13$, the mass loss dependence on metallicity 
can also be represented by

\begin{eqnarray}
\dot{M}~\propto~Z^m Z^{pq} & \propto~& Z^{0.85 + pq} \nonumber\\
                 \nonumber\\
       & &~{\rm for}~1/30~\le~Z/\zsun~\le~3
\label{eq_mdot2}
\end{eqnarray}
The overall result of these effects results in a dependence of mass loss $\dot{M} \propto Z^{0.69}$ for 
the O stars and in $\dot{M} \propto Z^{0.64}$ for the B supergiants.

These power-law dependencies derived with our Monte Carlo 
approach yield a stronger metallicity dependence than the value 
of $m$ = 1/2 that was derived by Kudritzki et al. (1987)
and has since been used in many evolutionary calculations (e.g. Langer 1991, Maeder 1992, 
Schaller et al. 1992, Meynet et al. 1994, Vassiliadis \& Wood 1994, Vanbeveren D. 1995, 
Iben et al. 1996, Deng et al. 1996).

\section{Complete mass-loss recipe}
\label{s_recipe}

In this section we present the ``complete'' theoretical mass loss formula for 
OB stars over the range in $\teff$ between 50~000 and 12~500 K and the 
range in $Z$ between 1/30 and 3 times $\zsun$. The mass-loss rate
as a function of {\it five} basic parameters will be provided. These parameters
are $M_*$, $L_*$, $\teff$, $\ratio$, and $Z$. 

First, some relationships for the bi-stability jumps have to
be connected. 
The position of this jump in $\teff$ now depends both on the metallicity $Z$ 
(this paper) and on the luminosity-to-mass ratio, i.e. $\Gamma_{\rm e}$ (Vink et al. 2000).
The characteristic density  $\langle\rho\rangle$ for the bi-stability jump around
$\teff \simeq$ 25~000 K can be determined by smoothly combining
Eq.~(\ref{eq_logZ}) from the present
paper with Eq.~(4) from Vink et al. (2000).
The joint result is given by

\begin{eqnarray}
{\rm log}< \rho > & = & - 14.94~(\pm 0.54) \nonumber\\
                  & &~+~0.85~(\pm 0.10)~{\rm log} (Z/\zsun) \nonumber\\
                  & &~+~3.2~(\pm 2.2)~\Gamma_e
\label{eq_charrho}
\end{eqnarray}
The positions (in $\teff$) of the several bi-stability jumps can now be found
using Eqs.~(\ref{eq_jump}),~(\ref{eq_jump2}),~(\ref{eq_jump3}) and (\ref{eq_charrho}).

We will divide our mass-loss recipe into two parts, taking into account
only the bi-stability jump around 25~000 K, since this jump is present 
at {\it all} metallicities in {\it all} panels of Fig.~\ref{f_mdotL5}.

If one wants a mass-loss rate for relatively high metallicity, say $Z/\zsun \ga 1$, for
low temperatures, $\teff \la$ 15~000 K, one should take into account the presence 
of the Fe {\sc iii}/{\sc ii} jump, and follow the strategy that was described in 
Vink et al. (2000). One may simply use Eq.~(\ref{eq_Bfit}; below) 
below the Fe {\sc iii}/{\sc ii} jump, but one should 
increase the constant by a factor of five (or $\Delta$ log  $\dot{M}$ = 0.70)
to a value of -5.99. The recipe can then be used until the point in the 
Hertzsprung-Russell Diagram (HRD) where line driven winds become 
inefficient (see Achmad \& Lamers 1997). We suggest that below the Fe {\sc iii}/{\sc ii} 
jump $\ratio$ = 0.7 (Lamers et al. 1995) is adopted.

If one needs a mass-loss rate for low metallicity, say $Z/\zsun \la 1/30$, at high
temperatures $\teff \ga$ 35~000 K, one should be aware of the carbon jump and a similar strategy
may be followed. 
Note that this jump is only present for cases where the wind density is 
weak, i.e. for stars with a relatively low luminosity. One can decrease 
the constant in Eq. (\ref{eq_Ofit}; below) by a factor of five (or $\Delta$ log $\dot{M}$ = 0.70)
to a value of -7.40. In case one does not know the value for $\vinf$ such as is the 
case for evolutionary calculations, one would like to know the appropriate change in terminal velocity
at the low $Z$ jump. Leitherer et al. (1992) have calculated the dependence of $\vinf$ on $Z$ and 
have found that $\vinf \propto Z^{0.13}$. Such a trend with metallicity has been confirmed by observations in 
the Magellanic Clouds, however, what happens to $\ratio$ at extremely low $Z$ is still an open question.
We stress that if the observed values for $\vinf$ at very low $Z$ turn out to be very different 
from the Galactic values, our mass-loss predictions can simply be scaled to accommodate the 
proper values of $\ratio$ and our recipe will still yield the corresponding mass-loss rates.

Now we can present the complete mass-loss recipe including the 
metallicity dependence. This can be done by simply adding the constant $Z$ dependence from 
Eq.~(\ref{eq_mdot2}) to the multiple linear regression relations from the Vink et al. (2000) 
recipe. We are indeed allowed to do so, as the $\dot{M}(Z)$ dependence was found to
be independent of other investigated stellar parameters (see Sect.~\ref{s_metal}).
For the hot side of the bi-stability jump $\sim 25~000$ K, the complete recipe is 
given by:

\begin{eqnarray}
{\rm log}~\dot{M} & = &~-~6.697~(\pm 0.061) \nonumber \\
                  & &~+~2.194~(\pm 0.021)~{\rm log}(L_*/{10^5}) \nonumber \\
                  & &~-~1.313~(\pm 0.046)~{\rm log}(M_*/30) \nonumber\\
                  & &~-~1.226~(\pm 0.037)~{\rm log}\left(\frac{\ratio}{2.0}\right) \nonumber \\
                  & &~+~0.933~(\pm 0.064)~{\rm log}(\teff/40 000) \nonumber\\
                  & &~-~10.92~(\pm 0.90)~\{{\rm log}(\teff/40 000)\}^{2} \nonumber\\
                  & &~+~0.85~(\pm 0.10)~{\rm log}(Z/\zsun) \nonumber\\
                  \nonumber\\
                  & &~{\rm for}~27~500 < \teff \le 50~000 {\rm K}
\label{eq_Ofit}
\end{eqnarray}
where $\dot{M}$ is in $\msun$ ${\rm yr}^{-1}$, $L_*$ and $M_*$ 
are in solar units and $\teff$ is in Kelvin. In this range the Galactic ratio of $\ratio$ = 2.6. 
As was noted in Sect.~\ref{s_assumptions}, if the values for $\vinf$ at other $Z$ are different 
from these Galactic values, then the mass-loss rates can easily be scaled accordingly.
 
For the cool side of the bi-stability jump, the complete recipe is

\begin{eqnarray}
{\rm log}~\dot{M} & = &~-~6.688~(\pm 0.080) \nonumber \\
                  & &~+~2.210~(\pm 0.031)~{\rm log}(L_*/{10^5}) \nonumber \\
                  & &~-~1.339~(\pm 0.068)~{\rm log}(M_*/30) \nonumber\\
                  & &~-~1.601~(\pm 0.055)~{\rm log}\left(\frac{\ratio}{2.0}\right) \nonumber \\
                  & &~+~1.07~(\pm 0.10)~{\rm log}(\teff/20 000) \nonumber\\
                  & &~+~0.85~(\pm 0.10)~{\rm log}(Z/\zsun) \nonumber\\
                  \nonumber\\
                  & &~{\rm for}~12~500 \le \teff \le 22~500 {\rm K}
\label{eq_Bfit}                  
\end{eqnarray}
where again $\dot{M}$ is in $\msun$ ${\rm yr}^{-1}$, $L_*$ and $M_*$ 
are in solar units and $\teff$ is in Kelvin. In this range the Galactic ratio of $\ratio$ = 1.3. 
In the critical temperature range between 22~500 $\le \teff \le$ 27~500 K, either 
Eq.~(\ref{eq_Ofit}) or Eq.~(\ref{eq_Bfit}) should be used depending on the position 
of the bi-stability jump given by Eq.~(\ref{eq_jump}).
A computer routine to calculate mass loss as a function of stellar
parameters is publicly available\footnote{see:~~~~astro.ic.ac.uk/$\sim$jvink/}.


\section{Comparison between theoretical $\dot{M}$ and observations at subsolar $Z$}
\label{s_comp}

\begin{figure}
\centerline{\psfig{file=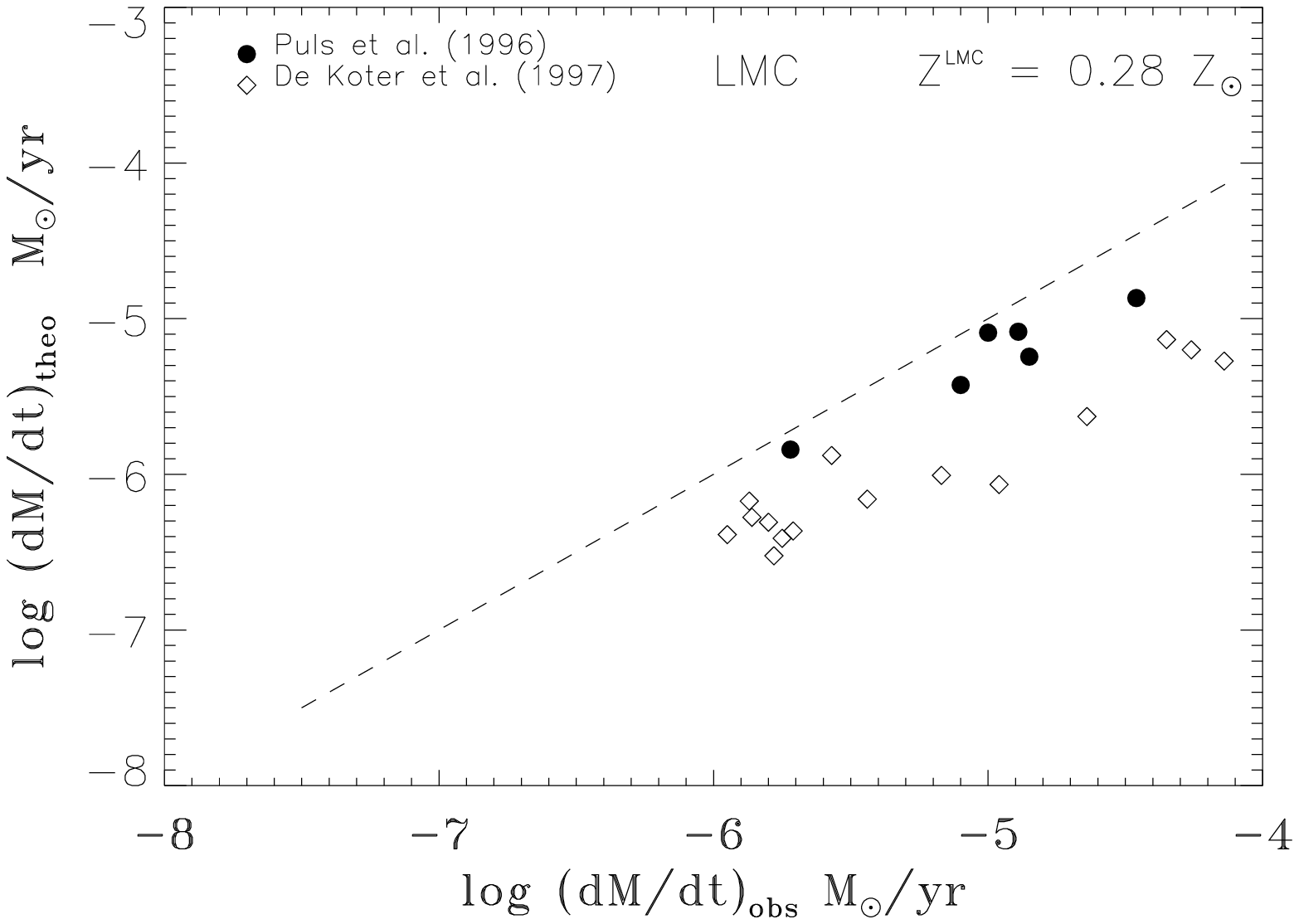, width = 10cm}}
\centerline{\psfig{file=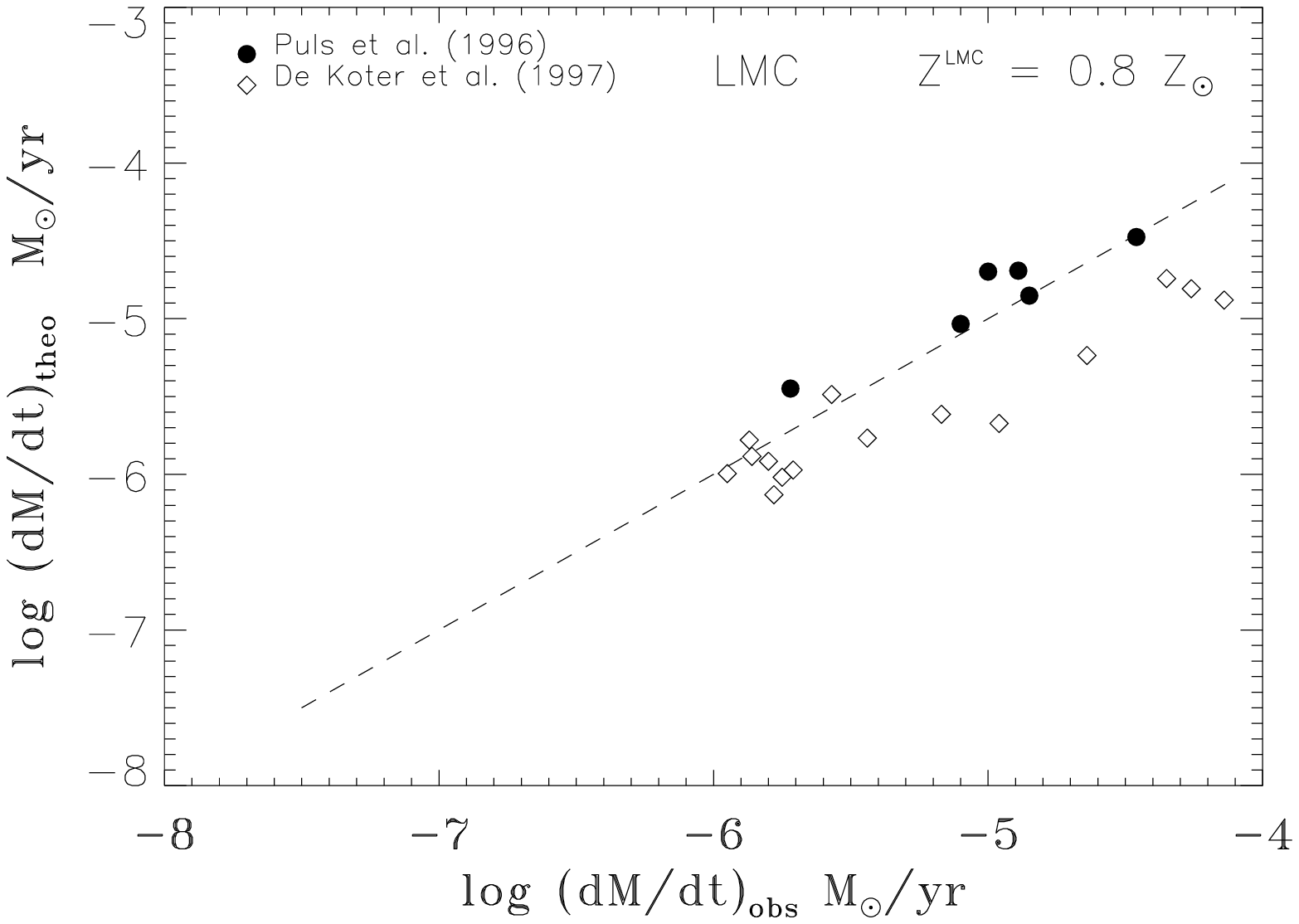, width = 10cm}}
\caption{Comparison between theoretical and observational $\dot{M}$ for O stars   
        in the LMC. The upper panel is for an adopted $Z_{\rm LMC}~=~0.28 \zsun$ and the lower panel
        is for an adopted metallicity $Z_{\rm LMC}~=~0.8 \zsun$.
        The Puls et al. (1996) H$\alpha$ rates and de de Koter rates 
        are indicated with different symbols.
        The dashed lines are one-to-one relations, tools for convenient comparison between
        observations and theory}
\label{f_zlmc}
\end{figure}

Now we will compare our mass-loss predictions for different $Z$ 
with the most reliable observational rates presently available.
Unfortunately, there are only substantial samples available in the literature 
for the relatively nearby Large Magellanic Cloud (LMC) and the Small Magellanic Cloud (SMC).
The metallicity of the LMC is only slightly smaller than the Galactic one and 
its absolute value is not accurately known.
What complicates a meaningful comparison is that there are differences in the observed 
stellar and nebular abundances. Additionally, there are abundance gradients 
present in these galaxies, which makes a good comparison between our predicted $\dot{M}(Z)$ 
dependence and the observed mass-loss rates of the LMC sample
rather difficult. As the metallicity difference between the Galaxy and 
the SMC is significantly larger, we should be able to test our predictions in a more meaningful
way with the observed rates of the SMC sample.

Following Kudritzki et al. (1987), we did not adopt the individual 
abundance patterns quoted for the Clouds (e.g. Dufour 1984). Instead we simply 
scaled down all abundances by a constant factor adopting:

\begin{eqnarray}
Z_{\rm LMC}~=~0.28~\zsun \nonumber\\    
Z_{\rm SMC}~=~0.10~\zsun
\end{eqnarray}
We are aware that the differential metal abundances in the Clouds could be different 
from the Galaxy due to a different stellar evolution at lower $Z$. However, 
we expect these effects to be of relatively minor importance, since the 
mass-loss rate at these metallicities ($Z \ga 1/10~\zsun$) is still mainly determined 
by iron.

The upper panel of Fig.~\ref{f_zlmc} shows the comparison between the 
observed LMC mass-loss rates and the theoretical values from our mass-loss recipe.
The scatter between observations and theory can be attributed to errors in 
the stellar parameters and the mass-loss determinations, but may also be 
due to differential metal abundance patterns in the LMC. Note that there 
is a systematic difference between the two sets of mass-loss determinations 
themselves (Puls et al. 1996 vs. de Koter et al. 1997, 1998). 
The possible systematic differences between these two sets have been discussed in 
de Koter et al. (1998). Nevertheless, {\it both} samples show an offset with
respect to our predictions. This could in principle be due to systematic errors 
in our predictions. However, since there 
is good agreement between observations and our predictions for a large sample of 
Galactic supergiants (Vink et al. 2000), we do not expect this to be the case. 
Perhaps the systematic offset is due to a too low assumed $Z$ for the LMC.
Haser et al. (1998) analysed individual O stars in the LMC and found metallicities 
significantly higher for these stars than usually derived from nebular abundance studies. 
Adopting the Haser et al. value of $Z = 0.8 \zsun$ derived for the LMC O star
SK-67$^o$166, for the whole LMC sample, 
there is much better agreement between our 
predictions and the observed mass-loss rates (see the lower panel in Fig.~\ref{f_zlmc}).
The scatter between observational and theoretical mass-loss rates decreases from 0.65 dex 
(1 $\sigma$) for the upper panel of Fig.~\ref{f_zlmc} to only 0.36 dex for the lower panel
of the figure.

\begin{figure}
\centerline{\psfig{file=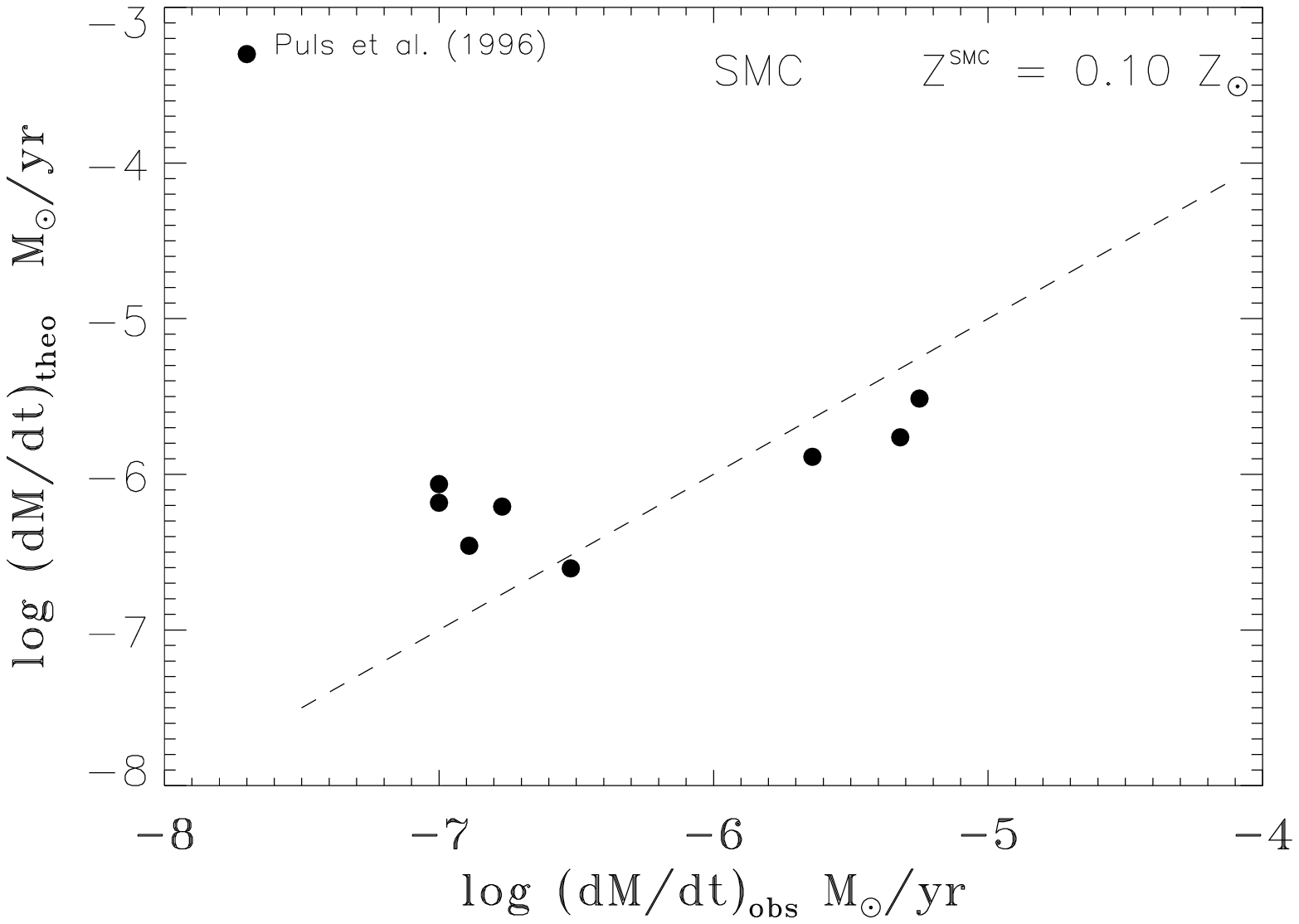, width = 10cm}}
\caption{Comparison between theoretical and observational $\dot{M}$ for O stars in  
        in the SMC with the adopted abundance of $Z_{\rm SMC}~=~0.10 \zsun$.
        The dashed line is the one-to-one relation, a tool for convenient 
        comparison between observations and theory.}
\label{f_zsmc}
\end{figure}
 
Figure~\ref{f_zsmc} shows the comparison between observed mass-loss rates and our
predictions for the sample of the SMC stars. The figure shows a reasonable agreement
between predictions and observations. 
We admit that there is quite a large scatter (0.55 dex) for which 
there may be several reasons. The important point at this stage is that 
the comparison with the SMC data yields good {\it average} agreement and thus 
yields support to the reliability of our mass loss recipe at metallicities 
other than solar.

For a test of our mass-loss recipe at extremely low $Z$, say $Z/\zsun <$  1/10, we 
will have to await new {\it Hubble Space Telescope (HST)} 
observations of some relatively nearby low metallicity galaxies. 


\section{Summary \& Conclusions}
\label{s_concl}

We have presented predictions of mass-loss rates for O and B stars 
over a wide range of metallicities. The calculations 
take the important effect of multiple line scattering into 
account in a consistent manner, using a ``Unified Monte Carlo approach''. It is shown that there 
is a {\it constant} universal metallicity dependence over 
a wide range of metal abundance, which can be represented by $\dot{M} \propto Z^{0.85} \vinf^p$, 
but that one needs to take into account some specific positions in the HRD where recombinations 
of Fe or CNO ions may cause the 
mass loss to increase dramatically and produce ``bi-stability'' jumps. 
It will be a challenge for the future to test our mass-loss 
recipe at extremely low $Z$ in local starbursting galaxies, where the 
difference in mass-loss rate compared to the solar neighbourhood can 
be significant. 

We can summarise the main results of the paper as follows:

\begin{enumerate}

\item{} We have calculated a grid of wind models and
mass-loss rates for a wide range of metallicities, covering 
$1/100 \le Z/\zsun \le 10$.

\item{} We have found that the mass loss vs. metallicity dependence
behaves as a power-law with $\dot{M}~\propto Z^{0.69}$ for O stars and $\dot{M}~\propto Z^{0.64}$ for 
B supergiants. This is in contrast to an often applied square-root dependence of mass loss on $Z$.

\item{} Although the $\dot{M}(Z)$ reaction is a constant function of $Z$, one 
should be aware of the presence of bi-stability jumps, where
the character of the line driving changes drastically due to 
a change in the wind ionization resulting in jumps in mass loss.
We have investigated the physical origins of these jumps and 
derived formulae that connect mass loss recipes at opposite 
sides of such bi-stability jumps. Additionally, we have made 
a distinction between the metal abundance derived from iron and from oxygen lines,
since observations of different galaxies have shown that the [Fe/O] abundance 
ratio varies with metallicity.

\item{} As our mass-loss predictions are successful in explaining 
the observed mass-loss rates for Galactic and Small Magellanic Cloud 
(Fig.~\ref{f_zsmc})
O-type stars, as well as in predicting the observed Galactic 
bi-stability jump, we believe that they are reliable 
and suggest that our mass-loss recipe be used in future 
evolutionary calculations, also at different $Z$.
A computer routine to calculate mass loss is publicly 
available at the address astro.ic.ac.uk/$\sim$jvink/.

\end{enumerate}


\begin{acknowledgements}

We thank the referee, Jon Bjorkman, for constructive comments that helped 
improve the paper. JV acknowledges financial support from the NWO Council for Physical 
Sciences. AdK acknowledges support from NWO Pionier grant 600-78-333 to L.B.F.M.
Waters and from NWO Spinoza grant 08-0 to E.P.J. van den Heuvel.

\end{acknowledgements}


\begin{thebibliography}{}

\bibitem[Abbott 1982]{abbott82}
       Abbott D.C., 1982, ApJ 259, 282

\bibitem[Abbott \& Lucy 1985]{abbott85}
       Abbott D.C., Lucy L.B., 1985, ApJ 288, 679

\bibitem[Achmad et al. 1997]{achmad97}
       Achmad L., Lamers H.J.G.L.M., Pasquini L., 1997, A\&A 320, 196

\bibitem[Allen 1973]{achmad73}
       Allen C.W., 1973, Astrophysical quantities, University of London, Athlone Press

\bibitem[Anders \& Grevesse 1989]{anders89}
       Anders E., Grevesse N., 1989, Geochim. Cosmochim. Acta 53, 197

\bibitem[Artymowicz 1993]{arty93}
        Artymowicz P., 1993, PASP 105, 1032 

\bibitem[Audouze 1987]{audou87}
       Audouze J., 1987, Observational Cosmology, IAU Symp. 124, 
       eds. A. Hewitt et al., Reidel Publ. p. 89

\bibitem[Castor et al. 1975]{castor75}
       Castor J.I., Abbott D.C., Klein R.I., 1975, ApJ 195, 157

\bibitem[Cen \& Ostriker 1999]{cen99}
       Cen R., Ostriker J.P., 1999, ApJ 519, 109

\bibitem[de Koter et al. 1993]{dekoter93}
       de Koter A., Schmutz W., Lamers H.J.G.L.M., 1993, A\&A 277, 561

\bibitem[de Koter et al. 1996]{dekoter96}
        de Koter A., Lamers H.J.G.L.M., Schmutz W., 1996, A\&A 306, 501

\bibitem[de Koter et al. 1997]{dekoter97}
       de Koter A., Heap S.R., Hubeny I., 1997, ApJ 477, 792

\bibitem[de Koter et al. 1998]{dekoter98}
       de Koter A., Heap S.R., Hubeny I., 1998,
       ApJ 509, 879

\bibitem[Deng et al. 1996]{deng96}
       Deng L., Bressan A., Chiosi C., 1996, A\&A 313, 145

\bibitem[Dufour 1984]{dufour84}
         Dufour R., 1984, IAU Symp. 108, p. 353

\bibitem[Garmany \& Conti 1985]{garm85}
        Garmany C.D., Conti P.S., 1985, ApJ 293, 407

\bibitem[Gayley 1995]{gay95}
         Gayley K.G., 1995, ApJ 454, 410

\bibitem[Hamann 1997]{hama97}
        Hamann F., 1997, ApJ 109, 279

\bibitem[Haser et al. 1998]{haser98}
        Haser S.M., Pauldrach A.W.A., Lennon D.J., 1998,  A\&A 330, 285

\bibitem[Iben et al. 1996]{iben96}
       Iben I.Jr., Tutukov A.V., Yungelson L.R., 1996, ApJ 456, 750

\bibitem[Kudritzki et al. 1987]{kudritzki87}
       Kudritzki R.-P., Pauldrach A.W.A., Puls J., 1987, A\&A 173, 293

\bibitem[Kudritzki et al. 1989]{kudritzki89}
       Kudritzki R.-P., Pauldrach A.W.A., Puls J., Abbott D.C., 1989, 
       AAP 219, 205

\bibitem[Kurucz 1988]{kurucz88}
       Kurucz R.L., 1988, IAU Trans., 20b, 168

\bibitem[Izotov \& Thuan]{izo99}
       Izotov Y.I., Thuan T.X., 1999, ApJ 511, 639

\bibitem[Lamers \& Leitherer]{lamers93}
        Lamers H.J.G.L.M., Leitherer C., 1993, ApJ 412, 771

\bibitem[Lamers et al. 1995]{lamers95}
        Lamers H.J.G.L.M., Snow T.P., Lindholm D.M., 1995, ApJ 455, 269

\bibitem[Langer 1991]{langer91}
        Langer N., 1991, A\&A 252, 669

\bibitem[Leitherer et al. 1989]{leith89}
        Leitherer C., Schmutz W., Abbott D.C., Hamann W.R., Wessolowski U., 1989, ApJ 346, 919

\bibitem[Leitherer et al. 1992]{leith92}
        Leitherer C., Robert C., Drissen L.,1992, ApJ 401, 596
        
\bibitem[Lucy \& Abbott 1993]{lucy93}
        Lucy L.B., Abbott D.C., 1993, ApJ 405, 738

\bibitem[Maeder 1992]{maeder92}
        Maeder A., 1992, A\&A 264, 105

\bibitem[Meynet et al. 1994]{meynet94}
        Meynet G., Maeder A., Schaller G., Schearer D., Charbonel C., 1994, A\&AS 103, 97

\bibitem[Pagel et al. 1992]{pagel92}
       Pagel B.E.J. Simonson E.A., Terlevich R.J., Edmunds M.G., 1992, MNRAS 255, 325 

\bibitem[Pagel \& Tautvaisiene 1995]{pagel95}
       Pagel B.E.J., Tautvaisiene G., 1995, MNRAS 276, 505

\bibitem[Pauldrach et al. 1986]{paul86}
        Pauldrach A.W.A., Puls J., Kudritzki R.P., 1986, A\&A 164, 86

\bibitem[Prinja 1987]{prinja87}
        Prinja R., 1987, MNRAS 228, 173

\bibitem[Puls 1987]{puls87}
        Puls J., 1987, A\&A 184, 227

\bibitem[Puls et al. 1996]{puls96}
        Puls J., Kudritzki R.P., Herrero A., et al., 1996, A\&A 305, 171

\bibitem[Puls et al. 2000]{puls00}
        Puls J., Springmann U., Lennon M., 2000, A\&AS 141, 23

\bibitem[Sargent \& Searle]{sarg70}
        Sargent W.L.W., Searle L., 1970, ApJL, 162, 155

\bibitem[Schaller et al., 1992]{schal92}
        Schaller G., Schaerer D., Meynet G., Maeder A., 1992, A\&AS 96, 269

\bibitem[Schmutz 1991]{schm91}
        Schmutz W., 1991, in: ``Stellar Atmospheres: Beyond Classical Models'',
        eds. Crivellari L., Hubeny I., Hummer D.G., NATO ASI Series C, Vol. 341, 191

\bibitem[Vanbeveren 1995]{vanb95}
        Vanbeveren D., 1995, A\&A 294, 107

\bibitem[Vassiliadis \& Wood 1994]{vass94}
        Vassiliadis E., Wood P.R., 1994, ApJS 92, 125

\bibitem[Vink 2000]{vink00a}
        Vink J.S., 2000, PhD thesis at Utrecht University

\bibitem[Vink et al. 1999]{vink99}
        Vink J.S., de Koter A., Lamers H.J.G.L.M., 1999, A\&A 350, 181

\bibitem[Vink et al. 2000]{vink00}
        Vink J.S., de Koter A., Lamers H.J.G.L.M., 2000, A\&A 362, 295

\end{thebibliography}
\end{document}